\documentclass{aa}  

\usepackage{graphicx}
\usepackage{txfonts}
\usepackage{subfigure}
\usepackage{threeparttable}
\bibpunct{(}{)}{;}{a}{}{,} 
\usepackage{gensymb}
\usepackage{natbib,twoopt}
\usepackage{footnote}
\usepackage[utf8]{inputenc}
\usepackage{array}
\usepackage{color}
\usepackage{mathtools}
\usepackage[export]{adjustbox}
\usepackage{amsmath,amssymb}
\usepackage{sidecap}
\sidecaptionvpos{figure}{c}
\begin{document} 
\def\mean#1{\left< #1 \right>}

   \title{X-ray study of the double radio relic Abell 3376 with Suzaku}

   \author{I. Urdampilleta \inst{1,2}
          \and H.Akamatsu \inst{1}
          \and F. Mernier \inst{1,3,4}
          \and J. S. Kaastra\inst{1,2}
          \and J. de Plaa\inst{1}
          \and T. Ohashi\inst{5}
          \and Y. Ishisaki\inst{5}
          \and H. Kawahara\inst{6,7}
             }
   \institute{SRON Netherlands Institute for Space Research, Sorbonnelaan 2, 3584 CA Utrecht, The Nether\-lands\\
              \email{i.urdampilleta@sron.nl}
         \and
            Leiden Observatory, Leiden University, PO Box 9513, 2300 RA Leiden, The Netherlands
         \and
         MTA-E\"otv\"os University Lend\"ulet Hot Universe Research Group, P\'azm\'any P\'eter s\'et\'any 1/A, Budapest, 1117, Hungary
         \and
      Institute of Physics, E\"otv\"os University, P\'azm\'any P\'eter s\'et\'any 1/A, Budapest, 1117, Hungary
       \and
             Department of Physics, Tokyo Metropolitan University, 1-1 Minami-Osawa, Hachioji, Tokyo 192-0397, Japan
          \and   
         Department of Earth and Planetary Science, The University of Tokyo, Tokyo 113-0033, Japan
         \and
Research Center for the Early Universe, School of Science, The University of Tokyo, Tokyo 113-0033, Japan\\}


 
  \abstract
   {We present an X-ray spectral analysis of the nearby double radio relic merging cluster Abell 3376 ($z$ = 0.046), observed with the \textit{Suzaku} XIS instrument. These deep ($\sim$360 ks) observations cover the entire double relic region in the outskirts of the cluster. These diffuse radio structures are amongst the largest and arc-shaped relics observed in combination with large-scale X-ray shocks in a merging cluster.  We confirm the presence of a stronger shock (${\cal M}_{\rm{W}}$ = 2.8~$\pm~0.4$)  in the western direction at $r\sim~$26$\arcmin$, derived from a temperature and surface brightness discontinuity across the radio relic. In the East, we detect a weaker shock (${\cal M}_{\rm{E}}$ = 1.5~$\pm~0.1$) at $r\sim~$8$\arcmin$, possibly associated to the 'notch' of eastern relic, and a cold front at $r\sim~$3$\arcmin$. Based on the shock speed calculated from the Mach numbers, we estimate that the dynamical age of the shock front is $\sim$0.6 Gyr after core passage, indicating that Abell 3376 is still an evolving merging cluster  and that the merger is taking place close to the plane of the sky. These results are consistent with simulations and optical and weak lensing studies from the literature.}

    


   \maketitle
%

\section{Introduction}

Galaxy clusters form hierarchically by the accretion and merging of the surrounding galaxy groups and subclusters. During these energetic processes, the intracluster medium (ICM) becomes turbulent and produces cold fronts, the interface between the infalling cool dense gas core of the subcluster and the hot cluster atmosphere, and shock waves, which propagate into the intracluster medium of the subclusters  \citep{Markevitch2007}.  Part of the kinetic energy involved in the merger is converted into thermal energy by driving these large-scale shocks and turbulence, and the other part is transformed into non-thermal energy. Shocks are thought to (re)accelerate electrons from the thermal distribution up to relativistic energies  by the first-order Fermi diffusive shock acceleration mechanism (hereafter DSA, \citealt{Bell1987,Blandford1987}). The accelerated electrons, in the presence of a magnetic field, may produce radio relics via synchrotron radio emission \citep{Ferrari2008,Feretti2012,Brunetti2014}. They are generally Mpc-scale sized, elongated and steep-spectrum radio structures \citep{Bruggen2012,Brunetti2014}, which appear to be associated to the shock fronts at the outskirts of merging clusters \citep{Finoguenov2010,Ogrean2013}.

Nowadays the discoveries of shocks coinciding with radio relics are increasing  \citep{Markevitch2005,Finoguenov2010,Macario2011,Ogrean2013,Bourdin2013,Eckert2016b, Sarazin2016, Akamatsu2013a,Akamatsu2015,Akamatsu2017}. However, not all the merging clusters present the same spatial distribution of the X-ray and radio components  \citep{Ogrean2014,Shimwell2016}. The study of the radial distribution of these thermal and non-thermal components allows to estimate the dynamical stage of the cluster as well as to understand how the shock propagates and heats the ICM. It can
also determine the physical association between radio relics
and shocks. The lack of connection between radio relics and shocks in several merging clusters, together with the low efficiency of DSA for ${\cal M}$~$\sim$~2--3 of these shocks, seems to suggest that in some cases the DSA assumption is not enough for the electron acceleration from the thermal pool \citep{Vink2014, Vazza2014,Skillman2013,Pinzke2013}. For this reason,  alternative mechanisms have been proposed recently, as for example the re-acceleration of pre-existing cosmic ray electrons \citep{Markevitch2005,Kang2012,Fujita2015,Fujita2016,Kang2017} and shock drift acceleration \citep{Guo2014a,Guo2014b,Guo2017}.  However, it is not clear what type of mechanism is involved and whether all relics need additional mechanisms or not. Some merging clusters present 
different results, as for example 'El Gordo' \citep{Botteon2016b} is consistent with the DSA mechanism. However, A3667 South \citep{Storm2017}, CIZA J2242.8+5301 \citep{Hoang2017} and A3411-3412 \citep{VanWeeren2017} challenge the direct shock acceleration of electrons from the thermal pool. Therefore, it is not possible to provide a definitive explanation for these discrepancies.

\begin{figure*}[!htb]
  
        \centering
        \includegraphics[width=1.0\textwidth]{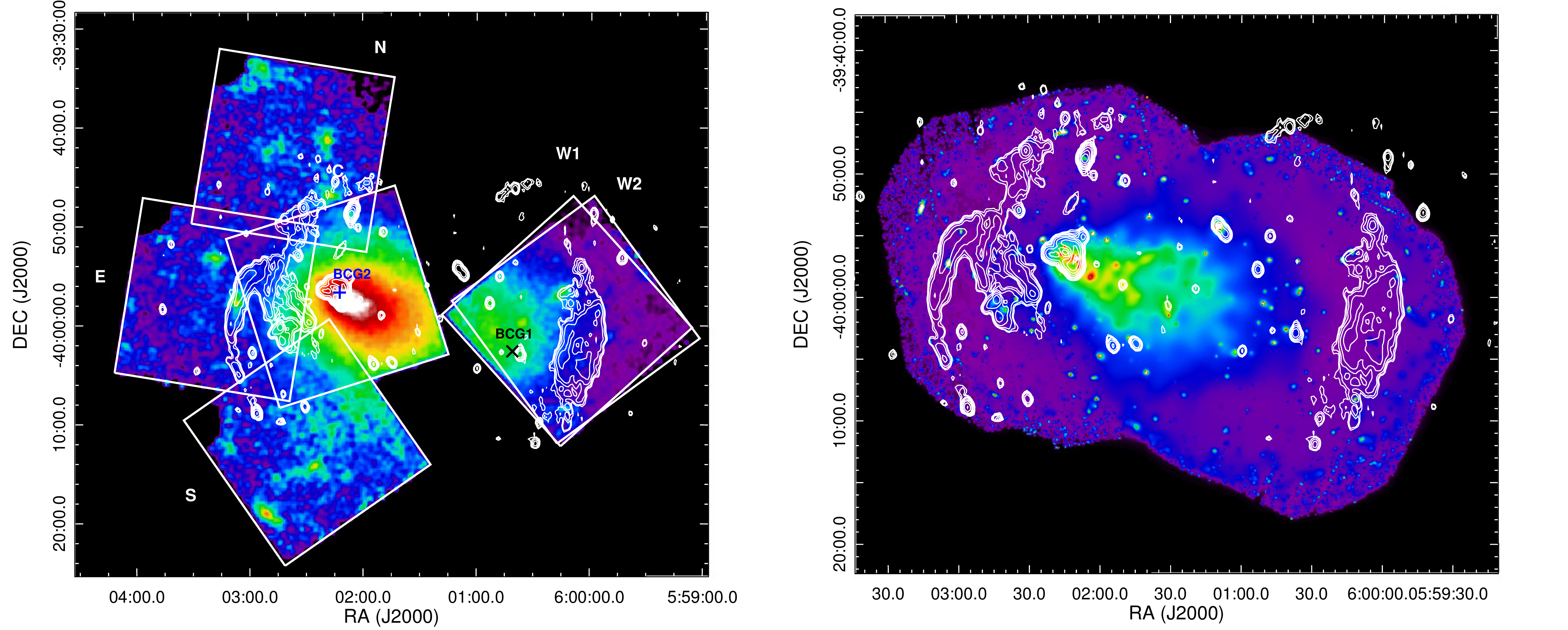}
        \label{fig:tworoundvarymujnegative}
    
    \caption{$Left$: \textit{Suzaku} smoothed image in the 0.5--10 keV band of A3376 (pixel = 8\arcsec\ and Gausssian 2-D smooth $\sigma$ = 16\arcsec). The white boxes represent the \textit{Suzaku} XIS FOVs for the C, W1, W2, N, E and S observations listed in Table \ref{tab:tab1}.  BCG1 is shown as a black cross and BCG2 as a blue cross. $Right$: \textit{XMM-Newton} image in the 0.5--10 keV band of A3376. The white contours correspond to VLA radio observations kindly provided by Dr. R. Kale.}
    \label{fig:fig1}
\end{figure*}

 \begin{table*}[!htb]
 \begin{center}
  \caption{Observations and exposure times.}
 \label{tab:tab1}
 \begin{threeparttable}
\begin{tabular}{ccccccc}
\hline
\hline
 \noalign{\smallskip}
Observatory & Name & Sequence ID &Position (J2000)&Observation&Exp.\tnote{a}&Exp.\tnote{b}\\
&&&(RA, Dec.)&starting date&(ks)&(ks)\\
 \noalign{\smallskip}
\hline
\noalign{\smallskip}
 \textit{Suzaku}&C&100043010&(90.56,-39.94)&2005-10-06&110.4&76.8 \\
 &W1&800011010&(90.05,-39.98)&2005-11-07&119.4&99.4 \\
 &W2&800011020&(90.04,-39.98)&2006-10-26&56.8&44.1 \\
 &N\tnote{c}&808029010&(90.64,-39.70)&2013-09-18&51.4&33.9 \\
 &E\tnote{c}&808028010&(90.81,-39.95)&2013-09-14&71.7&51.4\\
 &S\tnote{c}&808030010&(90.61,-40.18)&2013-10-15&61.3&44.5 \\
 &Q0551-3637&703036020&(88.19,-36.63)&2008-05-14&18.8&15.3 \\
 \noalign{\smallskip}
\hline
\noalign{\smallskip}
\textit{XMM-Newton}&XMM-E&0151900101&(90.54,-39.96)&2003-04-01&47.2&33.1 \\
&XMM-W&0504140101&(90.21,-40.03)&2007-08-24&46.1&40.8 \\
\noalign{\smallskip}
\hline
\noalign{\smallskip}
\textit{Chandra}&Chandra&3202&(90.54,-39.96)&2002-03-16&44.3&-- \\
\noalign{\smallskip}
\hline
\end{tabular}
\tiny
\begin{tablenotes}
    \item[a]\textit{Suzaku} data screening without COR2, \textit{XMM-Newton} and \textit{Chandra} total exposure time
    \item[b]\textit{Suzaku} data screening with COR2 > 8~GV; \textit{XMM-Newton} data screening  
    \item[c] Additional processing for XIS1
\end{tablenotes}
 \end{threeparttable}
 \end{center}
\end{table*}

  In this paper, we analyze the spatial distribution of thermal and non-thermal components in Abell 3376, (hereafter A3376), based on \textit{Suzaku} observations \citep{Mitsuda2007}.  We use in addition \textit{XMM-Newton} and \textit{Chandra} observations to support and confirm the results obtained with \textit{Suzaku}. A3376 is a nearby ($z$ = 0.046, \citealt{Struble1999}), bright and moderately massive ($M_{200}$~$\sim$~4--5~$\times$~10$^{14}$~M$_\odot$, \citealt{Durret2013,Monteiro-Oliveira2017}) merging galaxy cluster. This merging system has two giant ($\sim$Mpc) arc-shape  radio relics in the cluster outskirts, discovered by \cite{Bagchi2006}. The radio observations \citep{Bagchi2006,Kale2012,George2015}, see Fig. \ref{fig:fig1}, reveal complex radio structures at the western and eastern directions. In the West, A3376 shows a lowly polarized wide ($\sim$300 kpc) relic with a non-aligned magnetic field  \citep{Kale2012}. In the East, the relic is divided in three parts: the northern faint relic with steep spectral index; an elongated and polarized bright relic; and an additional 'notch' with a $\sim$500 kpc extension towards the center  ($\alpha\sim$ --1.5) \citep{Kale2012}. Radio spectral index studies, assuming diffusive shock acceleration, estimate ${\cal M}\sim$ 2--3, which is consistent with the previous X-ray study by \cite{Akamatsu2012b} of the western relic. A3376 includes two brightest cluster galaxies (BCG), coincident with two overdensities in projected galaxy distribution. BCG1  (ESO 307-13, $\rm{RA}=6^{\rm{h}} 00^{\rm{m}} 41\fs10$,
$\rm{Dec.}=-40\degr 02\arcmin 40\farcs 00$) belongs to the West subcluster, while BCG2  (2MASX J06020973-3956597, $\rm{RA}=6^{\rm{h}} 02^{\rm{m}} 09\fs70$,
$\rm{Dec.}=-39\degr 57\arcmin 05\farcs 00$) is located close to the X-ray peak emission at the East subcluster.   The N-body/SPH simulations of \cite{Machado2013} reproduce a bimodal merger system, with a mass ratio of 3--6:1, formed by one compact and dense subcluster, which crossed at high velocity and disrupted the core of the second massive subcluster $\sim$0.5--0.6 Gyr ago. This merger scenario was later corroborated by the optical analysis of \cite{Durret2013} and the weak lensing study of \cite{Monteiro-Oliveira2017}. 
  
  For this study, we use the values of protosolar abundances ($Z_\odot$) reported by \cite{Lodders2009}. The abundance of all metals are coupled to Fe. We use a Galactic absorption column of $N_{\rm{H,total}}$ = 5.6~$\times$~10$^{20}$ cm$^{-2}$ \citep{Willingale2013} for all the fits. We assume cosmological parameters $H_0$ = 70 km/s/Mpc, $\Omega_{\rm{M}}$ = 0.27 and $\Omega_\Lambda$= 0.73, respectively, which
give 54 kpc per 1 arcmin at $z$ = 0.046. The virial radius is adopted as $r_{200}$, the radius where the mean interior density is 200 times the critical density at the redshift of the source. For our cosmology and redshift we use the approximation of \cite{Henry2009}:
\begin{equation}
\label{eqn:eq0}
r_{200} = 2.77h^{-1}_{70}(\mean{kT}/10~\rm{keV})^{1/2}/\textit{E(z)}~\rm{Mpc},
\end{equation}

where $E(z)$=$(\Omega_{\rm{M}}(1 + z)^3 + 1 - \Omega_{\rm{M}})^{1/2}$, $r_{200}$ is 1.76 Mpc ($\sim$32.6\arcmin)  with $\mean{kT}$ = 4.2 keV  as calculated by \cite{Reiprich2013}.  All errors are given at 1$\sigma$ (68 $\%$) confidence level unless otherwise stated and all the spectral analysis made use of the modified Cash statistics \citep{Cash1979, Kaastra2017}.

  \section{Observations and data reduction}

  
  Table \ref{tab:tab1} and Fig. \ref{fig:fig1} summarize the observations used for the present analysis. As shown in Fig. 1 six different \textit{Suzaku}  pointings  have been used together with two additional \textit{XMM-Newton} observations. Three of the \textit{Suzaku} pointings are located in the central and western region (hereafter named as C, W1 and W2), which  were obtained in 2005 and 2006, respectively.  The other three observations (2013) cover the North (N), East (E) and South (S) of the outer eastern region. The combined observations cover the complete relic structures of A3376 as the western relic (W1 and W2), the center (C) and the eastern relic (N, E and S). The total effective exposure time of the \textit{Suzaku} data after screening and filtering of the cosmic-ray cut-off rigidity (COR2, \citealt{Tawa2008}) is  $\sim$360 ks. 

All observations have been performed with the normal 3x3 or 5x5 clocking mode of the X-ray imaging spectrometer (hereafter XIS, \citealt{Koyama2007}). This instrument had 4 active detectors: XIS0, XIS2 and XIS3 (front-side illuminated, FI) and XIS1 (back-side illuminated, BI). On November 9, 2006 the XIS2 detector suffered a micrometeorite strike\footnote{http://www.astro.isas.ac.jp/suzaku/doc/suzakumemo/suzakumemo-
2007-08.pdf} and was no longer operative. The XIS0 detector had a similar accident in 2010 and a part of the segment A was damaged\footnote{http://www.astro.isas.ac.jp/suzaku/doc/suzakumemo/suzakumemo-
2010-07v4.pdf}. In the  same year the amount of charge injection for XIS1 was increased from 2 to 6 keV, which leads to an increase in the non-X-ray background (NXB) level of XIS1. For that reason an additional screening has been applied to minimize the NXB level following the process described in the \textit{Suzaku} Data Reduction Guide \footnote{https://heasarc.gsfc.nasa.gov/docs/suzaku/analysis/abc/node8.html}.

 We used HEAsoft version 6.20 and CALDB 2016-01-28 for all \textit{Suzaku} analysis presented here. We have applied standard data screening with the elevation angle > 10 \degree\ above the Earth and cut-off rigidity (COR2) > 8 GV to increase the signal to noise. The energy range of 1.7-1.9 keV was ignored in the spectral fitting owing to the residual uncertainties present in the Si-K edge. 

For the point sources identification and exclusion, we used \textit{XMM-Newton} observations (ID: 0151900101 and 0504140101, see Table \ref{tab:tab1}). We applied the data reduction software SAS v14.0.0 for the \textit{XMM-Newton} EPIC instrument with the task named $edetect$\_$chain$. We used visual inspection for the point sources identification beyond the \textit{XMM-Newton} observations in the East. We derived the surface brightness (SB) profiles from the \textit{XMM-Newton} and \textit{Chandra} (ID: 3202, see Table \ref{tab:tab1}) observations. We used CIAO 4.8 with CALDB 4.7.6 for the data reduction of \textit{Chandra} observations.  Moreover, the \textit{Suzaku} offset observation of Q0551-3637, located at $\sim$4 degrees distance from A3376, was analyzed to estimate the sky background components as described in the following section.

\section{Spectral Analysis and results}

\subsection{Spectral analysis approach}

In our spectral analysis of A3376, we have assumed that the observed spectra include the following components: optically thin thermal plasma emission from the ICM, the cosmic X-ray background (CXB), the local hot bubble (LHB), the Milky Way halo (MWH) and non X-ray background (NXB). We have first  generated the non X-ray background spectra using \textit{xisnxbgen} \citep{Tawa2008}. Second, we have identified the point sources present in our data using the two observations of \textit{XMM-Newton} (see Table \ref{tab:tab1}) and extracted them with a radius of 1--3\arcmin\ from the \textit{Suzaku} observations. We have estimated the sky background emission consisting of CXB, LHB and MWH from the Q0551-3637 observation at 3.8\degree\ from A3376 (see Section 3.2). We assume that the sky background component is uniformly distributed along the A3376 extension. We have generated the redistribution matrix file (RMF) using \textit{xisrmfgen} and the ancillary response file (ARF) with \textit{xisimarfgen} \citep{Ishisaki2007} considering a uniform input image ($r$~=~20\arcmin).

 A proper estimation of the background components is crucial, specially in the outskirts of galaxy clusters, where usually the X-ray emission of the ICM is low and the spectrum can be background dominated. The X-ray shock fronts and radio relics are usually located at these outer regions.

 \begin{figure}[t!]
 \centering
\includegraphics[width=0.45\textwidth]{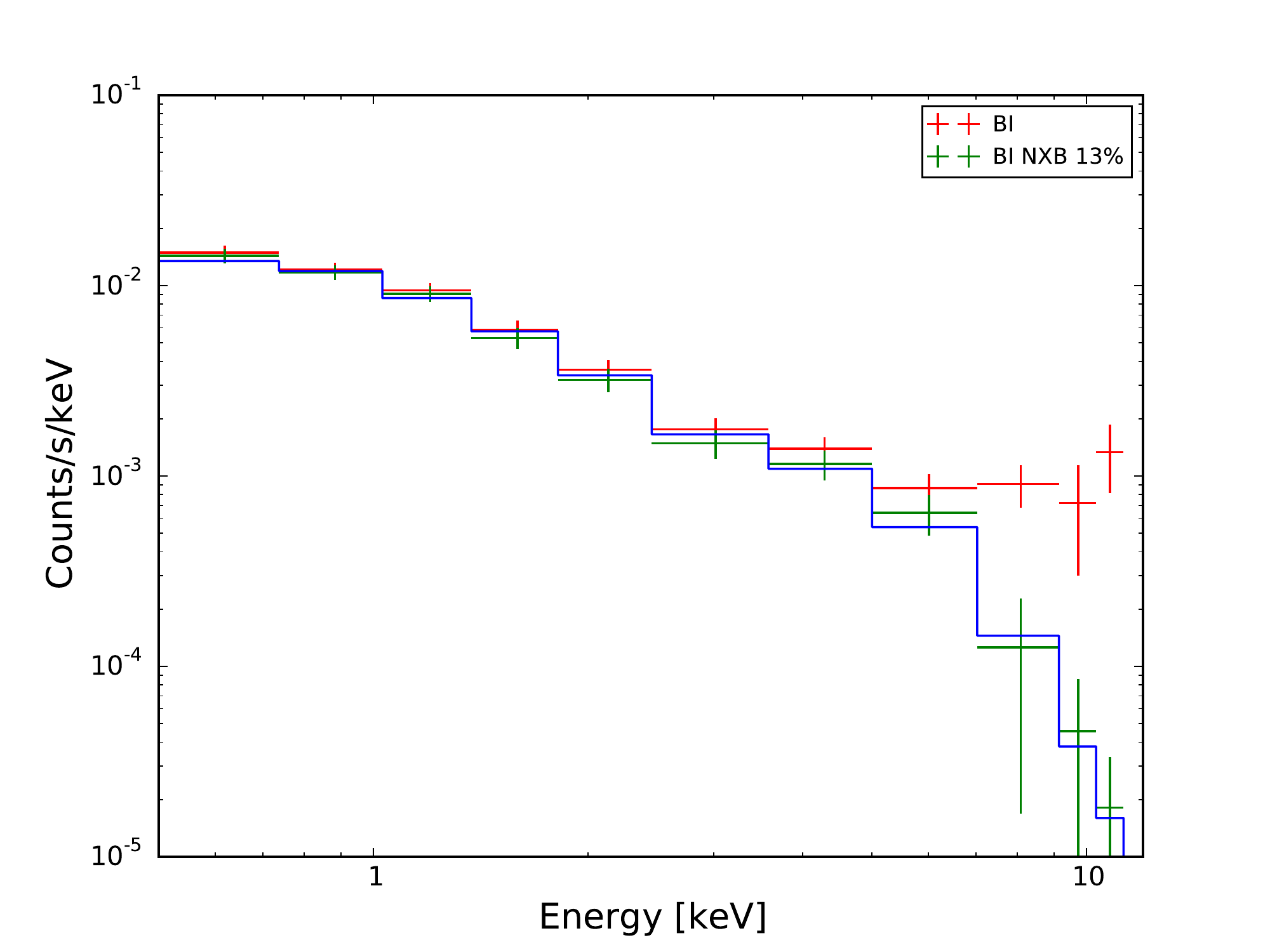}
\caption{Correction of the XIS1 excess in the outer region of the E observation of A3376. The best-fit ICM spectrum is shown as a blue line. The \textit{Suzaku} XIS BI spectra are shown without (red) and with (green) a 13\% increase in the NXB level, respectively. }
\label{fig:Fig3}
\end{figure}
\begin{figure}[ht!]
\includegraphics[width=0.5\textwidth]{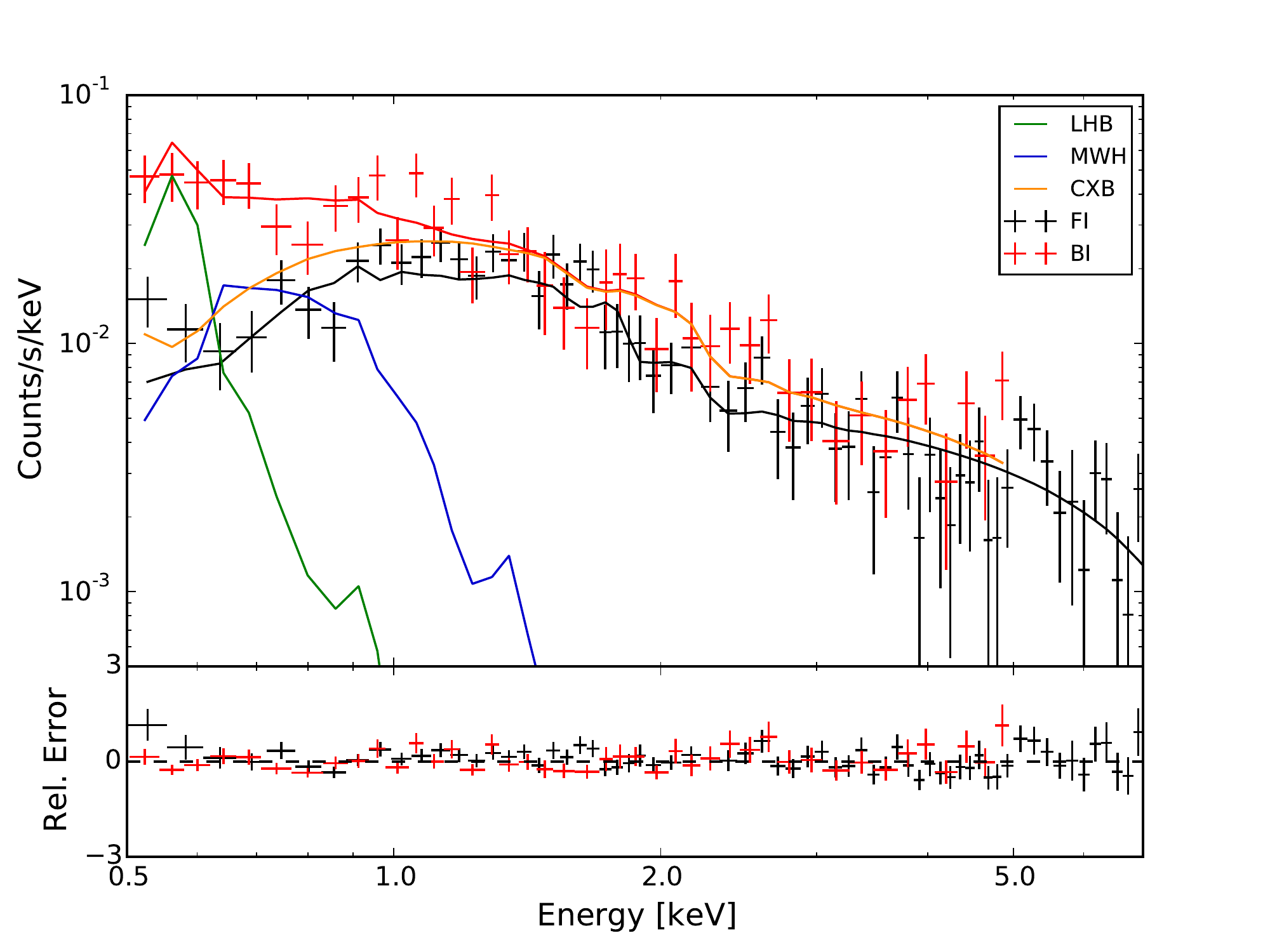}
\caption{The spectrum of the Q0551-3637 observation used for the sky background estimation. The \textit{Suzaku} XIS FI, BI, the LHB, MWH and CXB components are shown in black, red, green, blue and orange, respectively. The LHB, MWH and CXB has been represented relative to the BI spectrum.}
\label{fig:Fig2}
\end{figure}
\begin{table}[ht!]
 \begin{center}
  \caption{Background components derived from the Q0551-3637 observation.}
   \label{tab:tab2}
\begin{threeparttable}
 \begin{tabular}{cccc}
\hline
\hline
 \noalign{\smallskip}
 &LHB&MWH&CXB\\
  \noalign{\smallskip}
 \hline
 \noalign{\smallskip}
 \textit{Norm}\tnote{a}&108$~\pm~21$&2.9$~\pm~0.5$&4.9$~\pm~0.4$\\ 
  \noalign{\smallskip}
 \textit{kT} (keV)& 0.08 (fixed)&0.27 (fixed) & --\\
  \noalign{\smallskip}
 $\Gamma$&--&--& 1.41 (fixed)\\
 \noalign{\smallskip}
 \hline
  \noalign{\smallskip}
 C-stat/d.o.f.&& 130/103&\\
  \noalign{\smallskip}
  \hline

 \end{tabular}
 \tiny
 \begin{tablenotes}
    \item[a] For LHB and MWH norm in units of 10$^{71}$ m$^{-3}$ scaled by 400$\pi$\\
             For CXB norm in units of 10$^{51}$phs$^{-1}$keV$^{-1}$
\end{tablenotes}
  \end{threeparttable}
  \end{center}
 \end{table}

We performed a temperature deprojection assuming spherical symmetry to check the projection effect on the temperature profiles. We applied the method described in \cite{Blanton2003} for the post-shock regions, which are affected by projection of the emission of the outer and cooler regions. The results of the deprojection fits are consistent within 1$\sigma$ uncertainties with the projected temperatures. For this reason, we use the projected fits in this work.

 In our spectral analysis, we used SPEX\footnote{https://www.sron.nl/astrophysics-spex}  \citep{Kaastra1996} version 3.03.00 with SPEXACT version 2.07.00. We carried out the spectral fitting in different annular regions as detailed in the sections below. The spectra of the XIS FI and BI detectors were fitted simultaneously and binned using the method of optimal binning \citep{Kaastra2016}. For all the spectral fits, the NXB component is subtracted using the \textit{trafo} tool in SPEX.  The free parameters considered in this study are the temperature \textit{kT}, the normalization \textit{Norm} and the metal abundance Z for the inner regions $r$ $\leq$ 9\arcmin\ of the ICM component. For the outer regions ($r$ > 10\arcmin) we fixed the abundance to 0.3~$Z_\odot$ as explained by \cite{Fujita2008} and \cite{Urban2017}.

\subsection{Estimation of the background spectra} \label{bkg}
 The NXB component was obtained as the weighted sum of the XIS Night Earth observations and it was compiled for the same detector area and COR2 condition (COR2 > 8 GV) during 150 days before and after the observation date. In this way the long-term variation of the XIS detector background due to radiation damage can be constrained. The systematic errors corresponding to the NXB intensity are considered to be $\pm~3\%$  \citep{Tawa2008}.
For the most recent observations (E, N and S), where a new spectral analysis approach was proposed for XIS1\footnote{https://heasarc.gsfc.nasa.gov/docs/suzaku/analysis/abc/node8.html} because of the charge injection increase to 6 keV, we detected an excess of source counts in the high energy band (> 10 keV). We achieved to compensate this excess by increasing the NXB level by 13\%. In this way the source level (ICM) follows the spectral shape of the CXB at high energies (see Fig.  \ref{fig:Fig3}). 
\begin{figure*}[ht!]
    \centering
    \begin{minipage}{0.5\textwidth}
        \centering
        \includegraphics[width=1.0\textwidth]{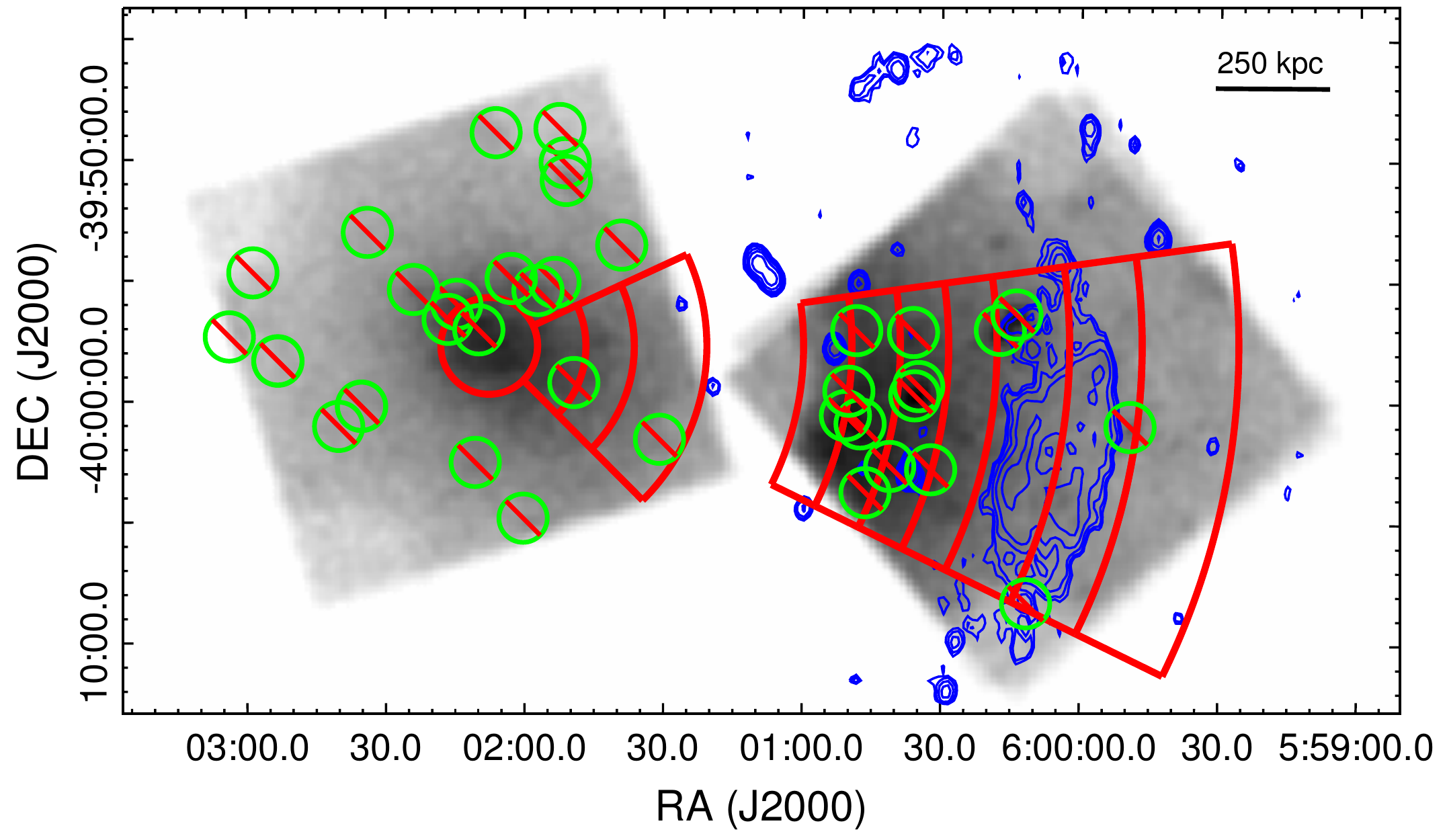} 
        \caption{A3376 Center and West observations in the band 0.5--10 keV. The blue contours represent VLA radio observations. The green circles represent the extracted points sources with  r=1\arcmin. The red annular regions are used for the spectral analysis detailed in Table \ref{tab:tab4}.}
        \label{fig:Fig4}
    \end{minipage}\hfill
    \begin{minipage}{0.48\textwidth}
        \centering
        \includegraphics[width=1.0\textwidth]{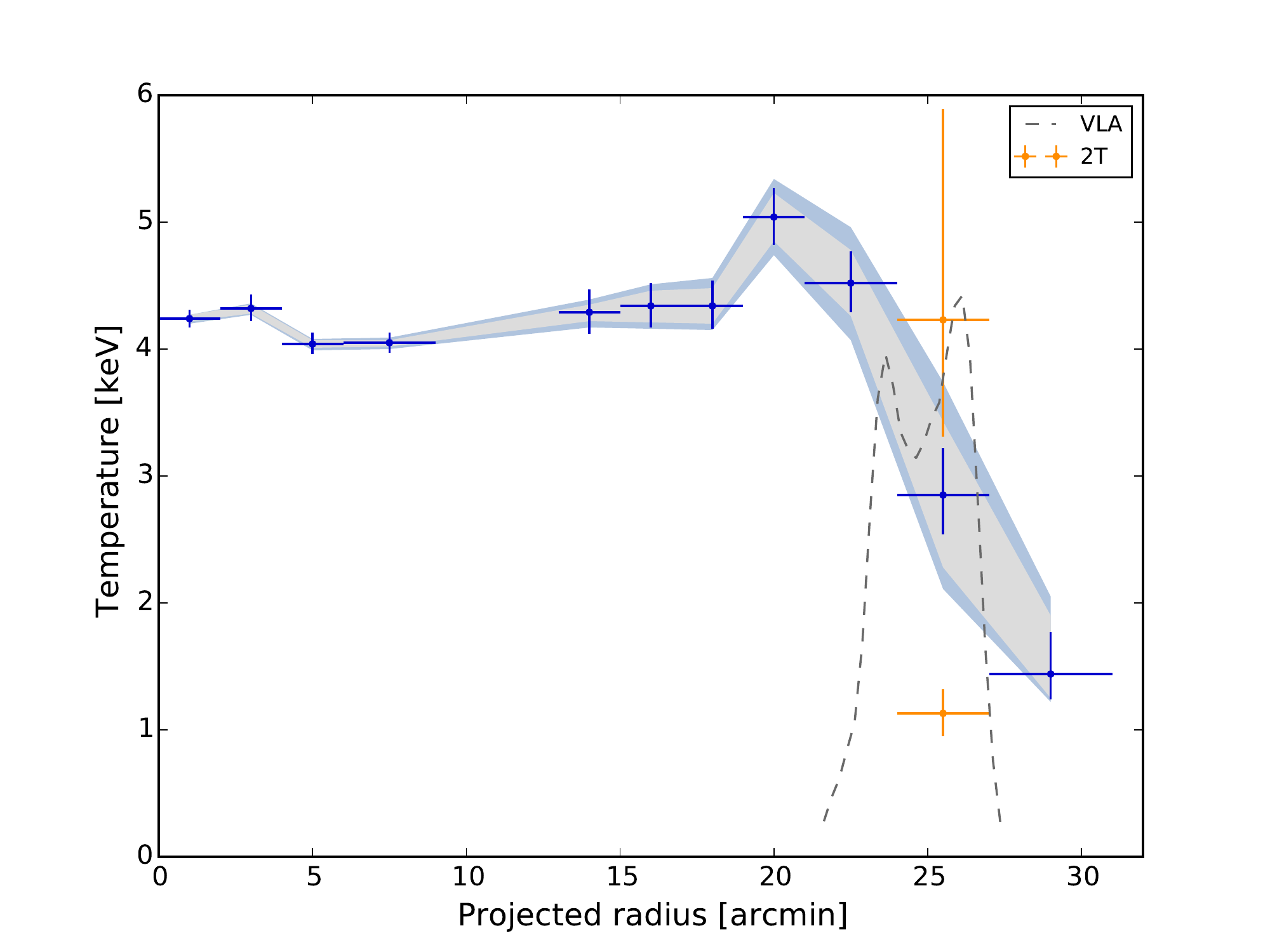} 
        \caption{The radial temperature profile for the western region. The gray and blue area represent the CXB and NXB systematic uncertainties. The orange points show the 2T model temperatures at the western radio relic. The dashed grey line is the VLA radio radial profile.}
        \label{fig:Fig5}
    \end{minipage}
\end{figure*}

\begin{table*}[ht!]
 \begin{center}  
 \caption{Best-fit parameters for the Center and West regions  shown in Fig. \ref{fig:Fig4}}.
   \label{tab:tab4}
 \begin{tabular}{cccccc}
 
\hline
\hline
\noalign{\smallskip}
 &Radius (\arcmin)&\textit{kT} (keV)&\textit{Norm} (10$^{73}$ m$^{-3}$)&$Z$ ($Z_\odot$)&C-stat/d.o.f.\\ 
 \noalign{\smallskip}
 \hline
\noalign{\smallskip}
 &1.0~$\pm~1.0$&4.27~$\pm~0.07$&7.92~$\pm~0.11$&0.53~$\pm~0.04$&803/793\\
 \noalign{\smallskip}
 Center&3.0~$\pm~1.0$&4.32~$\pm~0.11$&7.32$~\pm~0.16$&0.51~$\pm~0.06$&590/598\\
 \noalign{\smallskip}
 &5.0~$\pm~1.0$&4.04~$\pm~0.09$&6.32~$\pm~0.15$&0.41~$\pm~0.05$&634/606\\
 \noalign{\smallskip}
 &7.5~$\pm~1.5$&4.05~$\pm~0.08$&5.52~$\pm~0.09$&0.39~$\pm~0.04$&606/583\\
 \noalign{\smallskip}
 \hline
 \noalign{\smallskip}
 &14.0~$\pm~1.0$&4.29~$\pm~0.18$&2.68~$\pm~0.08$&0.3 (fixed)&327/337\\
 \noalign{\smallskip}
 &16.0~$\pm~1.0$&4.34~$\pm~0.18$&1.90~$\pm~0.06$&0.3 (fixed)&415/415\\
 \noalign{\smallskip}
 &18.0~$\pm~1.0$&4.34~$\pm~0.19$&1.41~$\pm~0.04$&0.3 (fixed)&451/448\\
 \noalign{\smallskip}
 West&20.0~$\pm~1.0$&5.04~$\pm~0.23$&0.90~$\pm~0.02$&0.3 (fixed)&690/656\\
  \noalign{\smallskip}
 &22.5~$\pm~1.5$&4.52~$\pm~0.24$&0.42~$\pm~0.01$&0.3 (fixed)&796/752\\
  \noalign{\smallskip}
 &25.5~$\pm~1.5$&2.85~$\pm~0.34$&0.15~$\pm~0.01$&0.3 (fixed)&670/576\\
  \noalign{\smallskip}
 &29.0~$\pm~2.0$&1.44~$\pm~0.27$&0.03~$\pm~0.01$&0.3 (fixed)&329/244\\
 \noalign{\smallskip}
  \hline

 \end{tabular}
  \end{center}
 \end{table*}
 
As mentioned in Sect. 2, the point source identification in the \textit{Suzaku} FOV was realized using two \textit{XMM-Newton} observations.  The SAS task named $edetect\_chain$ was applied in four different bands (0.3--2.0 keV, 2.0--4.5 keV, 4.5--7.5 keV, and 7.5--12 keV) using EPIC pn and MOS data. We used a simulated maximum likelihood value of 10. After this, we combined the detections in these four bands and estimated the flux of each source in a circle with a radius of 0.6\arcmin. We established the minimum of flux detection of the extracted sources as $S_{\rm{c}}$ = 10$^{-17}$ W m$^{-2}$ in the energy band 2--10 keV. Although this limit is lower than the level reported by \cite{Kushino2002}  $S_{\rm{c}}$ = 2~$\times$~10$^{-16}$ W m$^{-2}$,  we obtained an acceptable log$N$– log$S$ distribution contained within the \cite{Kushino2002} limits (see their Fig. 20). We excluded in the \textit{Suzaku} observations the identified point sources with a radius of 1 arcmin in order to account for the point spread function (PSF) of XIS  \citep{Serlemitsos2007}. As a result, our CXB intensity after the point sources extraction for the 2--10 keV band was estimated as 5.98~$\times$~10$^{-11}$ W m$^{-2}$ sr$^{-1}$. This value is within $\pm10\%$ agreement with \cite{Cappelluti2017}, \cite{Akamatsu2012b} and  \cite{Kushino2002}.

For sanity check, we compared the sky background level of ROSAT observations for the outermost region of A3376 ($r$~=~30\arcmin-60\arcmin) and the offset pointing Q0551-3637 ($r$~=~3\arcmin-20\arcmin) using the HEASARC X-ray Background Tool\footnote{https://heasarc.gsfc.nasa.gov/cgi-bin/Tools/xraybg/xraybg.pl} in the band R45 (0.4--1.2 keV). This band contains most of the emission of the sky background. The R45 intensities in the units of 10$^{-6}$ counts/s/arcmin$^2$  are 125.3~$\pm$~5.0 for the A3376 outer ring and  114.4~$\pm$~18.5 for Q0551-3637. Both values are in good agreement.

We used the same $S_{\rm{c}}$ = 10$^{-17}$ W m$^{-2}$  for extracting the point sources in the offset \textit{Suzaku} observation of Q0551-3637. We excluded as well a circle of 3\arcmin\ centered around the quasar with the same name, which is the  brightest source of this region. Thereafter, we fitted the resultant spectra using $N_{\rm{H}}$ = 3.2~$\times$~10$^{20}$ cm$^{-2}$ \citep{Willingale2013}, considering the emission of three sky background components at redshift zero and metal abundance of 1. We used for the FI and BI detectors the energy ranges of 0.5--7.0 keV and 0.5--5.0 keV, respectively. The two Galactic components are modelled with: LHB (fixed \textit{kT} = 0.08 keV), unabsorbed $cie$ model (collisional ionization equilibrium in SPEX) and MWH (fixed \textit{kT} = 0.27 keV), absorbed $cie$. The third component is the CXB modelled as an absorbed $powerlaw$ with fixed $\Gamma$ = 1.41. The  complete sky background model is: 
\begin{equation}
\label{eqn:eq1}
cie_{\rm{LHB}}+abs*(cie_{\rm{MWH}}+powerlaw_{\rm{CXB}}).
\end{equation}

 The best-fit parameters are listed in Table \ref{tab:tab2} and the sky background components are shown in Fig. \ref{fig:Fig2}. For the calculation of the systematic uncertainties of the CXB fluctuation due to unresolved point sources ($\sigma/I_{\rm{CXB}}$) we have used equation (3) proposed by \cite{Hoshino2010} in each spatial region of Fig. \ref{fig:Fig4} and Fig. \ref{fig:Fig7} as:
\begin{equation}
\label{eqn:eq10}
\frac{\sigma}{I_{\rm{CXB}}} = \frac{\sigma_{\rm{Ginga}}}{I_{\rm{CXB}}}~\bigg(\frac{\Omega_{\rm{e}}}{\Omega_{\rm{e,Ginga}}}\bigg)^{-0.5}~\bigg(\frac{S_{\rm{c}}}{S_{\rm{c,Ginga}}}\bigg)^{0.25}
\end{equation}
where ($\sigma_{\rm{Ginga}}/I_{\rm{CXB}}$)~=~5, $\Omega_{\rm{e,Ginga}}$~=~1.2 deg${^2}$,                         $S_{\rm{c,Ginga}}$~=~6~$\times$~10$^{-15}$ W m$^{-2}$,  $S_{\rm{c}}$ = 10$^{-17}$ W m$^{-2}$ and $\Omega_{\rm{e}}$ is the effective solid angle of the detector. This way,  we have included the effect of systematic uncertainties related to the NXB intensity  \citep[$\pm~3\%$,][]{Tawa2008} and the CXB fluctuation, which varies between 8 to 27$\%$. The contributions of these systematic uncertainties are shown in Fig. \ref{fig:Fig5} and Fig. \ref{fig:Fig8}.

 \subsection{Spectral analysis along the western region}

For the spectral analysis of the western region, we have analysed several annular regions centred on the X-ray emission peak centroid ($\rm{RA}=6^{\rm{h}} 02^{\rm{m}} 07\fs66$,
$\rm{Dec.}=-39\degr 57\arcmin 42\farcs 74$) once the point sources (see green circles in Fig. \ref{fig:Fig4}) have been excluded. The regions are a circle in the center with $r$~=~2$\arcmin$ and annuli between 2\arcmin--4\arcmin, 4\arcmin--6\arcmin, 6\arcmin--9\arcmin, 13\arcmin--15\arcmin, 15\arcmin--17\arcmin, 17\arcmin--19\arcmin, 19\arcmin--21\arcmin, 21\arcmin--24\arcmin, 24\arcmin--27\arcmin and 27\arcmin--31\arcmin. For all of them, the FI (used energy range 0.5--10 keV) and BI (0.5--7.0 keV) spectra have been fitted simultaneously. We have analyzed the NXB subtracted spectra with the normalization \textit{Norm},  temperature \textit{kT} and metal abundance $Z$ for the inner region $r$ $\leq$ 9\arcmin\ as free parameters.  The sky background components have been fixed to the values presented in Table \ref{tab:tab2}.

 \begin{figure}[t!]
  \centering
\includegraphics[width=0.45\textwidth]{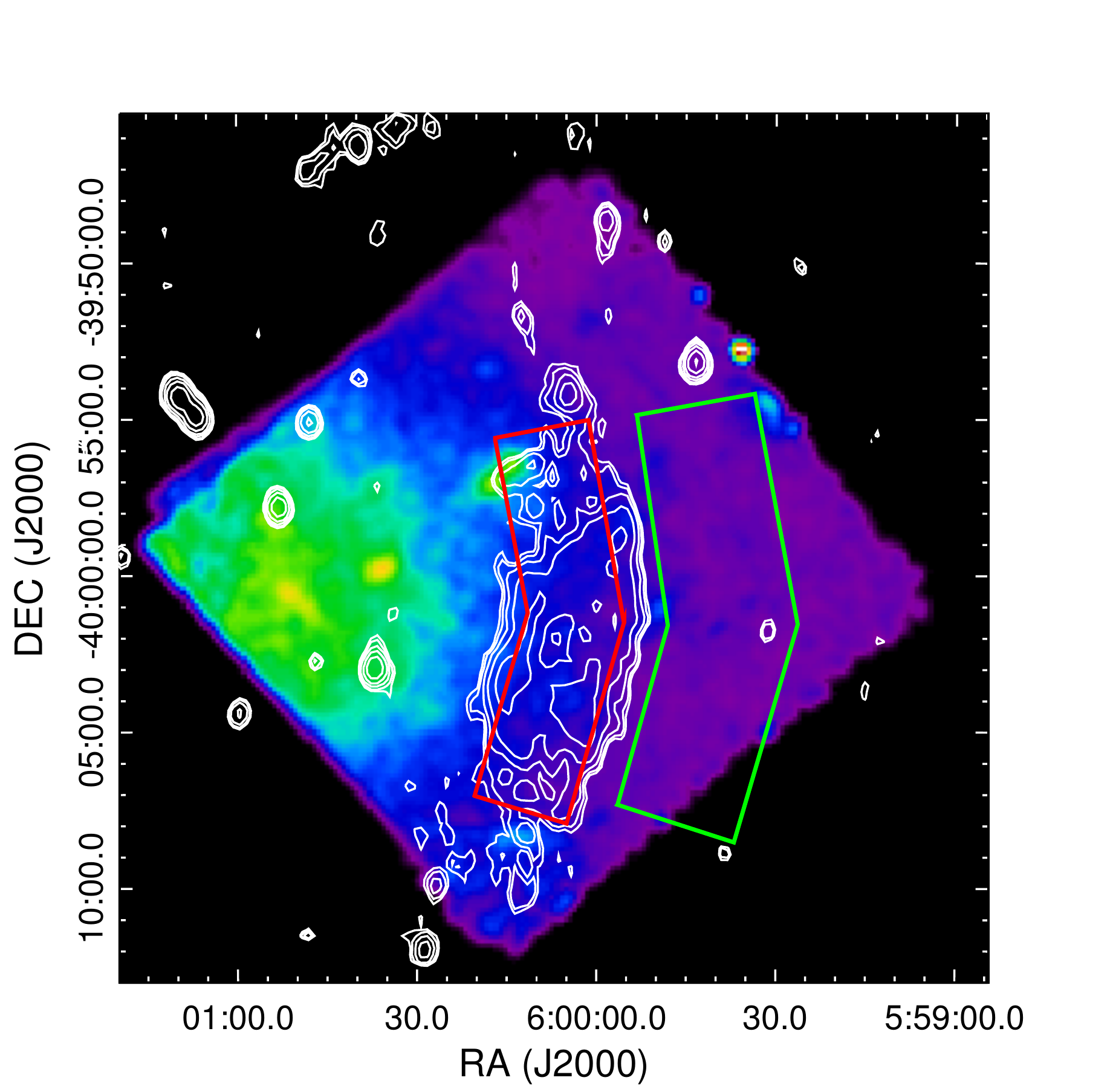}
\caption{A3376 W1 and W2 NXB subtracted images in the band 0.5-10 keV.  The white contours correspond to VLA radio observations. The green and red polygonal regions are the pre and post-shock regions of the western radio relic, respectively.}
\label{fig:Fig6}
\end{figure}
 
\begin{table}[t]
 \begin{center}
 \caption{Best-fit parameters for the pre- and post-shock regions at the western relic shown in Fig. \ref{fig:Fig6}}.
   \label{tab:tab5}
 \begin{tabular}{cccc}
  \hline
\hline
\noalign{\smallskip}
 &\textit{kT} (keV)&\textit{Norm} (10$^{71}$ m$^{-3}$)&C-stat/d.o.f.\\ 
 \noalign{\smallskip}
 \hline
 \noalign{\smallskip}
 Post&4.22~$\pm$~0.26&27.7~$\pm$~1.0 &680/654\\
 \noalign{\smallskip}
 Pre& 1.27~$\pm$~0.29&2.3~$\pm$~0.8 &638/516 \\
 \noalign{\smallskip}
 \hline
 \end{tabular}
  \end{center}
 \end{table}

 \begin{figure}[b!]
\centering 
  
\includegraphics[width=0.40\textwidth]{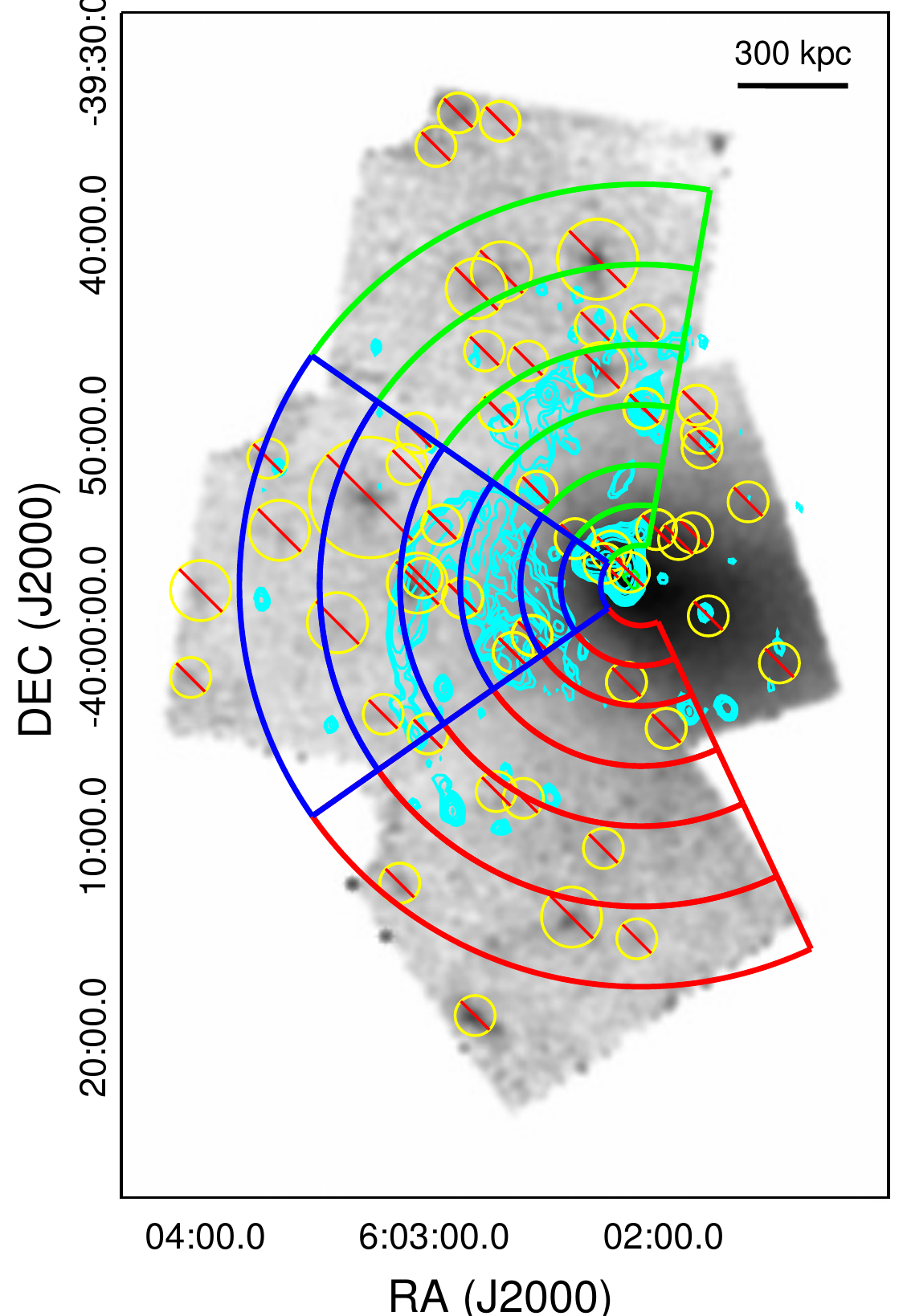}
 
\caption{A3376 Center, North, East and South images in the band 0.5--10 keV. The cyan contours represent VLA radio observations. The yellow circles represent the extracted points sources. The green (North), blue (East) and red (South) annular regions are used for the spectral analysis detailed in Table \ref{tab:tab7}, \ref{tab:tab8} and \ref{tab:tab9}, respectively.}
\label{fig:Fig7}
\end{figure}
\begin{figure}[b!]
\centering
\begin{minipage}[t]{1\textwidth}
\includegraphics[width=0.50\textwidth]{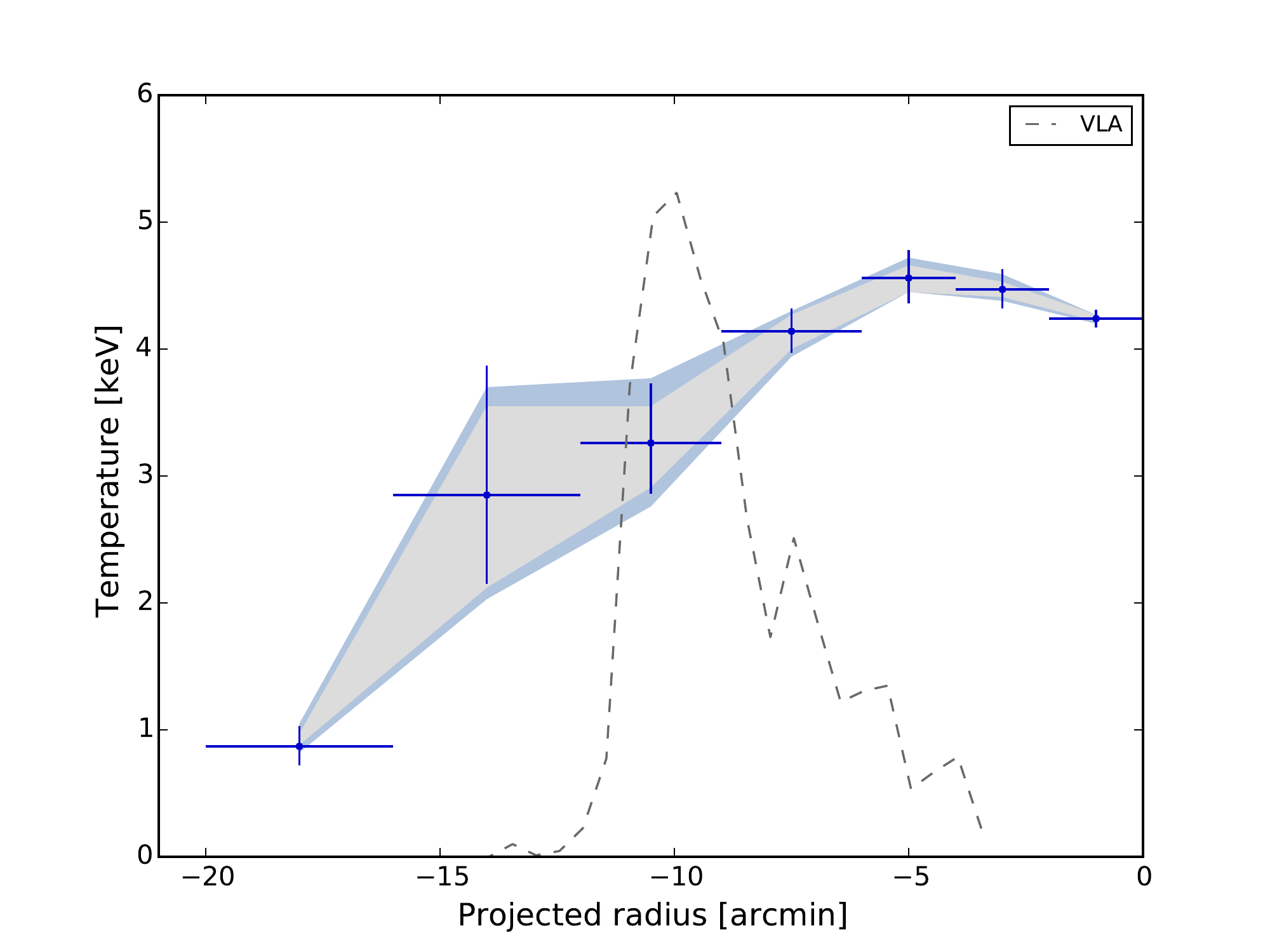}
\end{minipage}
\hfill
\begin{minipage}[t]{1\textwidth}
\includegraphics[width=0.50\textwidth]{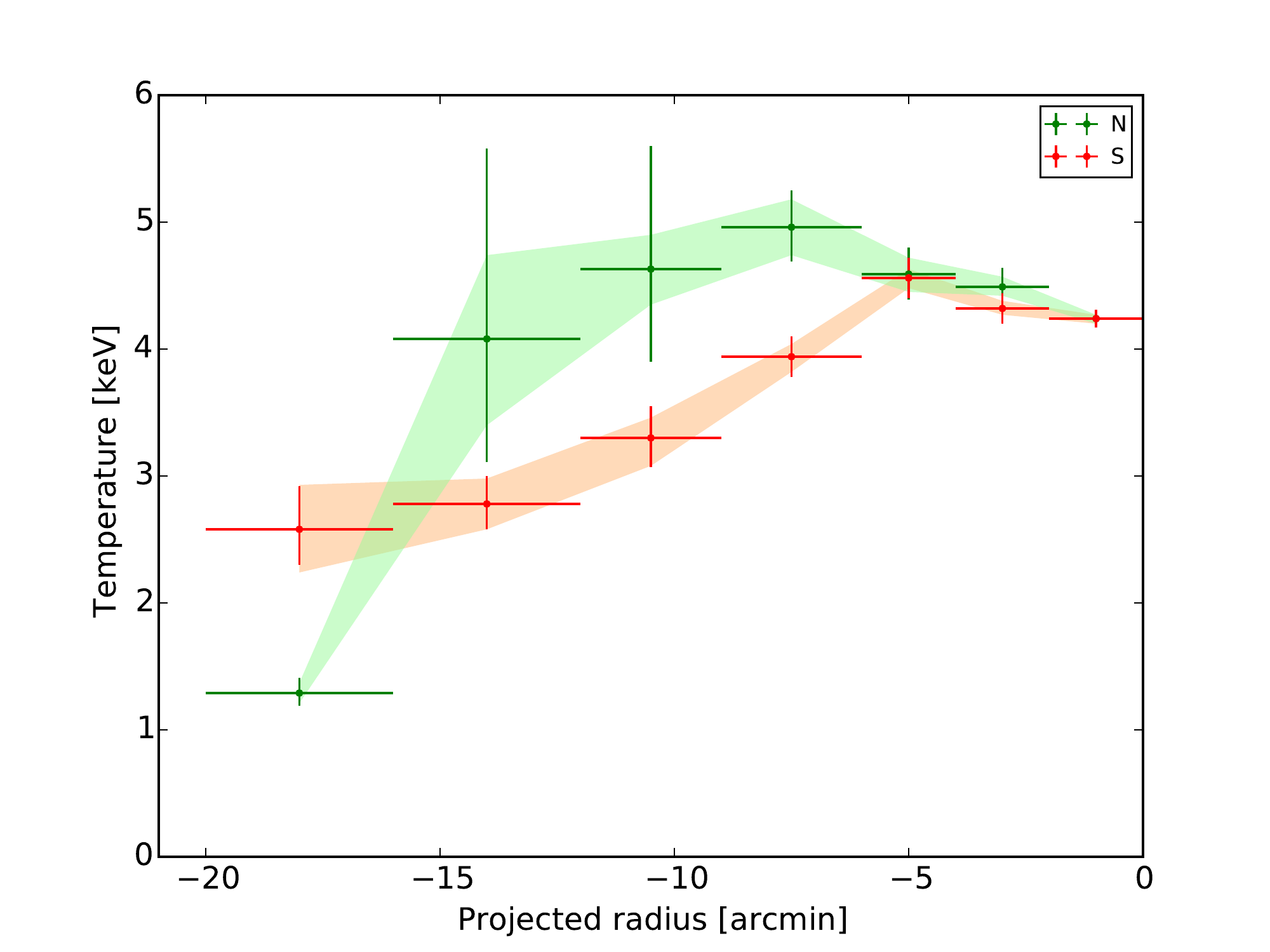}
\end{minipage}
\caption{Radial temperature profile for the eastern region (Top: East and Bottom: North and South). Top: Gray and blue areas represent the CXB and NXB systematic uncertainties. The dashed grey line is the VLA radio radial profile. Bottom: Green and red areas represent the CXB and NXB systematic uncertainties for North and South, respectively.  }
\setlength{\unitlength}{5cm}
\label{fig:Fig8}
\end{figure}

  The best-fit parameters  are summarized in Table \ref{tab:tab4}. In general, we obtained a good fit for all regions (C-stat/d.o.f. < 1.2). The resulting radial temperature profile is shown in Fig. \ref{fig:Fig5}. It includes in gray and blue the systematic uncertainties due to the CXB fluctuation and NXB, respectively, as mentioned above. The radial profile shows an average temperature of $\sim$4 keV in the central region of the cluster as found earlier by \cite{DeGrandi2002}, \cite{Kawano2008} and \cite{Akamatsu2012b}.  At $r$~=~20$\arcmin$ the temperature increases slightly to $\sim$5 keV and drops smoothly to $\sim$1.4 keV beyond the western radio relic. Although there is a temperature decrease, there is not a clear discontinuity in the temperature at the radio relic.  One possible explanation is that the annular region  at $r$~=~24\arcmin--27\arcmin\ contains gas from the pre and post relic region, which have two different temperatures. In order to investigate this aspect we included a second $cie$ model in this region. As a result, we obtained two distinct temperatures: $kT_{1}$~=~4.2~$\pm$~1.3 keV and $kT_{2}$~=~1.1~$\pm$~0.2 keV (see orange points in Fig. \ref{fig:Fig5}) which are consistent with adjacent regions indicating  the presence of multi-temperature structure within the extraction region. Therefore, this could be a preliminary indication of a temperature jump in the radio relic area and the possible presence of a shock front.

 \begin{figure*}[ht!]
  \centering
  \begin{minipage}[h]{0.40\textwidth}
    \centering
    \includegraphics[width=1\textwidth]{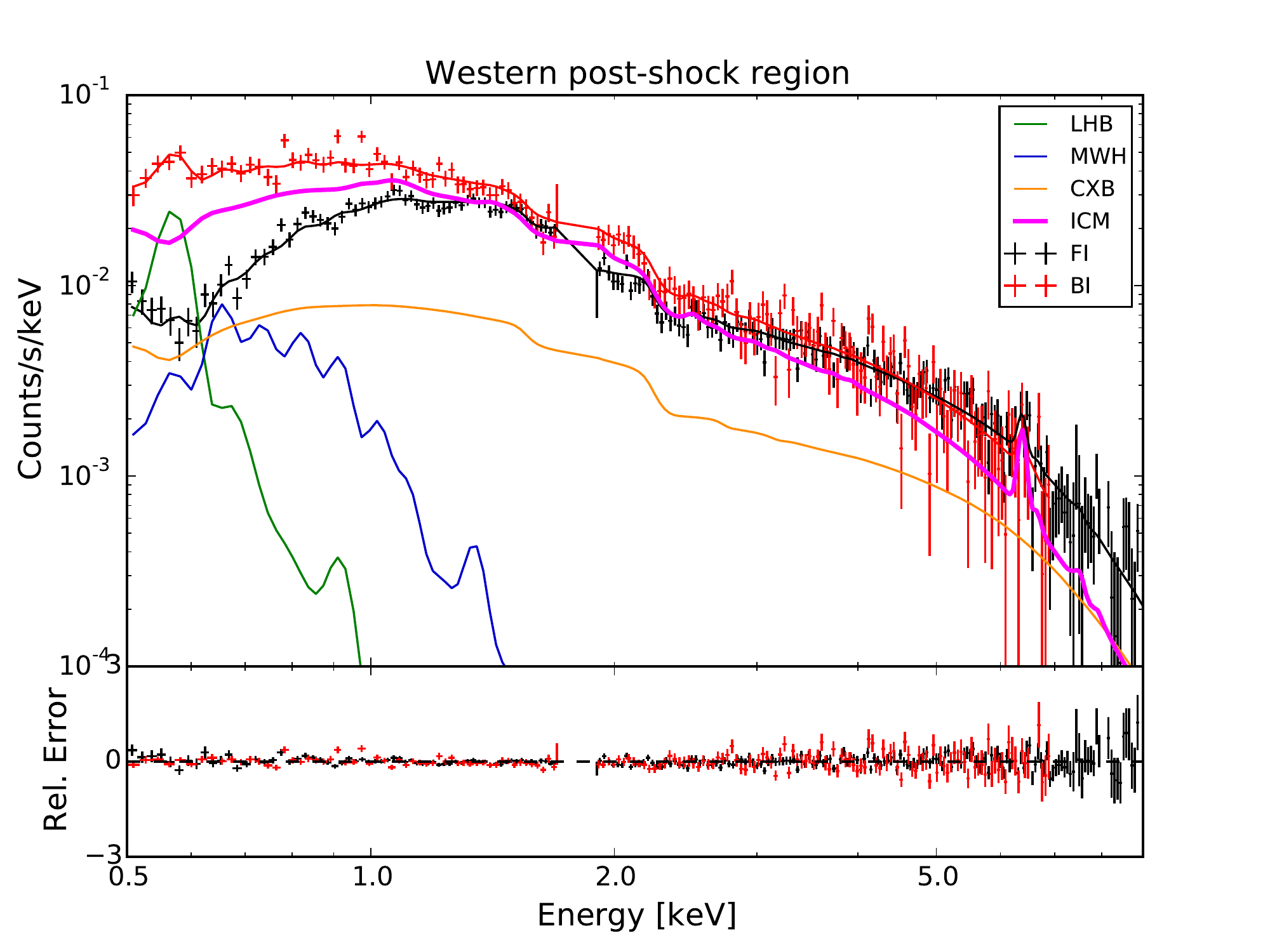}
  \end{minipage}
 \begin{minipage}[h]{0.40\textwidth}
    \centering
    \includegraphics[width=1\textwidth]{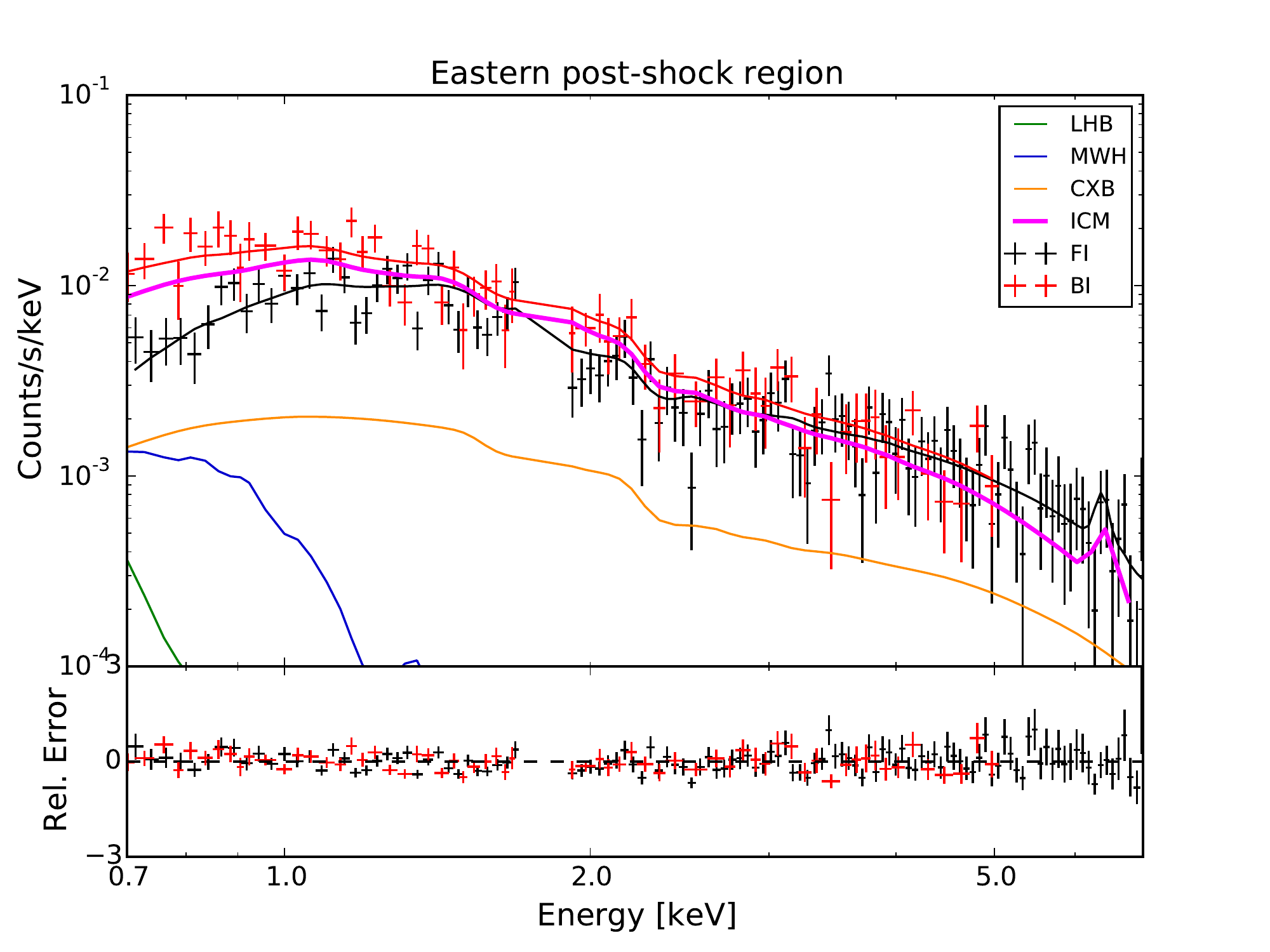} 
 \end{minipage}
\caption{Left:  NXB-subtracted spectrum of western post-shock region in the 0.5--10 keV band. Right: eastern post-shock spectrum in the 0.7--7.0 keV band. The FI (black) and BI (red) spectra are fitted with the ICM model together with the CXB and Galactic emission. The ICM is shown in magenta. The LHB, MWH and CXB are represented by green, blue and orange curves, respectively. The ICM,LHB, MWH and CXB has been represented relative to the BI spectrum.}
\label{fig:Fig14}
\end{figure*}
\begin{table*}[t!]
 \begin{center}
 \caption{Best-fit parameters for the Center and East regions shown in Fig. \ref{fig:Fig7}.}
   \label{tab:tab8}
 \begin{tabular}{cccccc}
 \hline
\hline
\noalign{\smallskip}
 &Radius (\arcmin)&$kT$ (keV)&\textit{Norm} (10$^{73}$ m$^{-3}$)&$Z$ ($Z_\odot$)&C-stat/d.o.f.\\ 
\noalign{\smallskip}
\hline
 \noalign{\smallskip}
 Center&3.0~$\pm~1.0$&4.47$~\pm~0.16$&4.11$~\pm~0.11$&0.33~$\pm~0.06$&491/451\\ 
  \noalign{\smallskip}
 &5.0~$\pm~1.0$&4.56$~\pm~0.21$&1.96~$\pm~0.06$&0.25~$\pm~0.07$&361/340\\ 
 \noalign{\smallskip}
 \hline
  \noalign{\smallskip}
 C\&E&7.5~$\pm~1.5$&4.14$~\pm~0.18$&0.95~$\pm~0.03$&0.34~$\pm~0.08$&661/630\\ 
  \noalign{\smallskip}
 \hline
  \noalign{\smallskip}
 &10.5~$\pm~1.5$&3.26$~\pm~0.44$&0.36~$\pm~0.02$&0.3 (fixed)&258/266\\ 
  \noalign{\smallskip}
 East&14.0~$\pm~2.0$&2.85$~\pm~0.86$&0.12~$\pm~0.02$&0.3 (fixed)&170/169\\ 
  \noalign{\smallskip}
 &18.0~$\pm~2.0$&0.87$~\pm~0.16$&0.02~$\pm~0.01$&0.3 (fixed)&180/196\\
 \noalign{\smallskip}
  \hline
 \end{tabular}
  \end{center}

 \end{table*}
\begin{table*}[t!]
 \begin{center}
 \caption{Best-fit parameters for the Center and North regions   shown in Fig. \ref{fig:Fig7}.}
   \label{tab:tab7}
 \begin{tabular}{cccccc}
 \hline
\hline
\noalign{\smallskip}
 &Radius (\arcmin)&$k$T (keV)&\textit{Norm} (10$^{73}$ m$^{-3}$)&$Z$  ($Z_\odot$)&C-stat/d.o.f.\\ 
 \noalign{\smallskip}
\hline
\noalign{\smallskip}
 &3.0~$\pm~1.0$&4.49$~\pm~0.15$&3.81$~\pm~0.09$&0.38~$\pm~0.06$&522/509\\ 
 \noalign{\smallskip}
 Center&5.0~$\pm~1.0$&4.59$~\pm~0.21$&1.73~$\pm~0.05$&0.40~$\pm~0.09$&331/363\\
 \noalign{\smallskip}
  &7.5~$\pm~1.5$&4.96$~\pm~0.28$&0.94~$\pm~0.03$&0.36~$\pm~0.11$&343/327\\
 \noalign{\smallskip}
 \hline
 \noalign{\smallskip}
 &10.5~$\pm~1.5$&4.63$~\pm~0.85$&0.48~$\pm~0.04$&0.3 (fixed)&129/118\\
 \noalign{\smallskip}
 North&14.0~$\pm~2.0$&4.08$~\pm~1.24$&0.17~$\pm~0.02$&0.3 (fixed)&126/157\\
 \noalign{\smallskip}
 &18.0~$\pm~2.0$&1.29$~\pm~0.11$&0.12~$\pm~0.02$&0.3 (fixed)&192/173\\
 \noalign{\smallskip}
  \hline
 \end{tabular}
  \end{center}

\end{table*}
\begin{table*}[t!]
 \begin{center}
 \caption{Best-fit parameters for the Center and South regions  shown in Fig. \ref{fig:Fig7}.}
   \label{tab:tab9}
 \begin{tabular}{cccccc}
 \hline
\hline
\noalign{\smallskip}
  &Radius (\arcmin)&\textit{kT} (keV)&\textit{Norm} (10$^{73}$ m$^{-3}$)&$Z$  ($Z_\odot$)&C-stat/d.o.f.\\ 
\noalign{\smallskip}
\hline
\noalign{\smallskip}
 &3.0~$\pm~1.0$&4.32$~\pm~0.12$&5.04$~\pm~0.11$&0.38~$\pm~0.06$&578/559\\
 \noalign{\smallskip}
 Center &5.0~$\pm~1.0$&4.56$~\pm~0.16$&3.18$~\pm~0.08$&0.47~$\pm~0.07$&396/399\\
 \noalign{\smallskip}
 &7.5~$\pm~1.0$&3.94$~\pm~0.16$&1.82~$\pm~0.05$&0.25$~\pm~0.07$&326/359\\
 \noalign{\smallskip}
 \hline
  \noalign{\smallskip}
 &10.5~$\pm~1.5$&3.30$~\pm~0.24$&0.66~$\pm~0.03$&0.3 (fixed)&386/370\\
 \noalign{\smallskip}
 South&14.0~$\pm~2.0$&2.78$~\pm~0.21$&0.44~$\pm~0.02$&0.3 (fixed)&271/251\\
 \noalign{\smallskip}
 &18.0~$\pm~2.0$&2.58$~\pm~0.31$&0.25~$\pm~0.02$&0.3 (fixed)&204/213\\
 \noalign{\smallskip}
  \hline

\end{tabular}
  \end{center}
 \end{table*}

 As a next step, we have defined two additional regions in order to obtain the temperature upstream and downstream of the possible X-ray shock (Fig. \ref{fig:Fig6}). We have introduced a separation of $\sim$1\arcmin\ between pre and post-shock region to avoid a possible photon leakage from the brighter region due to limited PSF of XRT \citep{Akamatsu2015,Akamatsu2017}. The pre-shock region is located   beyond the radio relic edge (the green polygonal region in Fig. \ref{fig:Fig6}) and the post-shock region is inside the radio relic edge (the red polygonal region in Fig. \ref{fig:Fig6}). The resulting best-fit parameters for these pre and post-shock regions are shown in Table \ref{tab:tab5} and the spectrum of the post-shock region is shown in Fig. \ref{fig:Fig14}. In this new analysis, the temperature shows a significant drop from $kT_{\rm{post}}$~=~4.22~$\pm$~0.26  keV to $kT_{\rm{pre}}$~=~1.27~$\pm$~0.29  keV. These ICM temperatures of the pre and post-shock regions agree with previous \textit{Suzaku} results \citep[$kT_{\rm{post}}~=~4.68^{+0.42}_{-0.24}$ to $kT_{\rm{pre}}$~=~1.34~$\pm$~0.42 keV,][]{Akamatsu2012b}. We discuss the shock properties related to the western radio relic in Sect. \ref{ssec:shockprop}.

\subsection{Spectral analysis along the eastern region}

\begin{figure}[ht!]
 \centering
\includegraphics[width=0.40\textwidth]{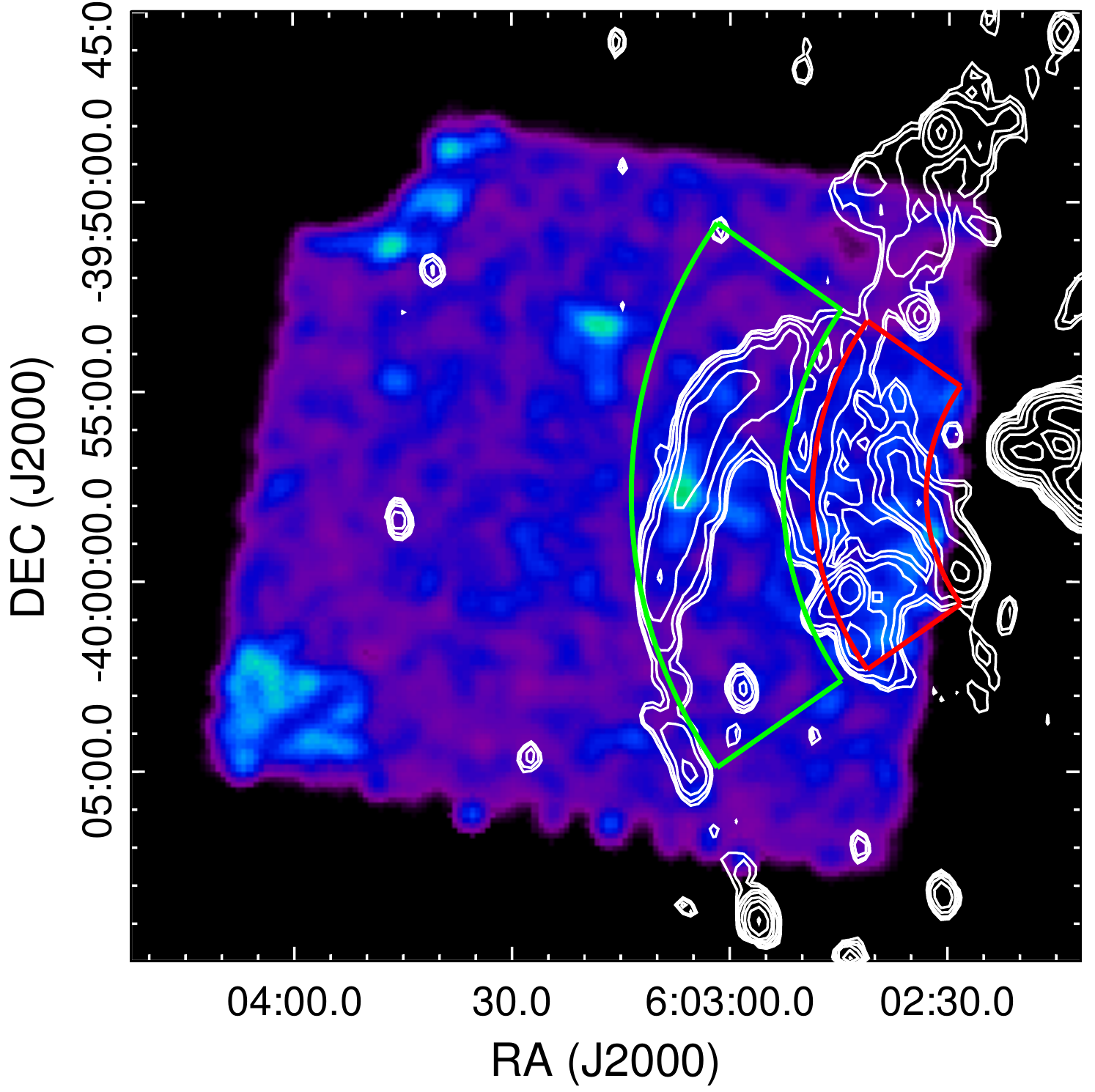}
\caption{A3376 East NXB subtracted images in the band 0.5--10 keV. The white contours correspond to VLA radio observations. The green and red annular regions are the pre and post-shock regions of the eastern and northern radio relic, respectively.}
\setlength{\unitlength}{5cm}
\label{fig:Fig9}
\end{figure}
\begin{table}[h!]
 \begin{center}
 \caption{Best-fit parameters for the pre- and post-shock regions at the East region   shown in Fig. \ref{fig:Fig9}.}
  \label{tab:tab6}
 \begin{tabular}{cccc}
  \hline
\hline
\noalign{\smallskip}
 &\textit{kT} (keV)&\textit{Norm} (10$^{72}$ m$^{-3}$)&C-stat/d.o.f.\\ 
 \noalign{\smallskip}
 \hline
 \noalign{\smallskip}
 Post&4.71~$\pm$~0.42&11.7~$\pm$~0.1 &192/187\\
 \noalign{\smallskip}
 Pre& 3.31~$\pm$~0.44&3.2~$\pm$~0.2 &182/173\\
 \noalign{\smallskip}
 \hline
 \end{tabular}
  \end{center}
 \end{table} 
 In the spectral analysis of the eastern region we have studied three different directions: N, E and S, corresponding to the observations with the same names (see Table \ref{tab:tab1}). We used annular regions, with the same centroid as the western region ($\rm{RA}=6^{\rm{h}} 02^{\rm{m}} 07\fs66$,
$\rm{Dec.}=-39\degr 57\arcmin 42\farcs 74$), between 2\arcmin--4\arcmin, 4\arcmin--6\arcmin, 6\arcmin--9\arcmin, 9\arcmin--12\arcmin, 12\arcmin--16\arcmin\ and 16\arcmin--20\arcmin. Fig. \ref{fig:Fig7} shows green regions for N, blue regions for E and red regions for S. We fitted the FI (0.7--7.0 keV) and BI (0.7--5.0 keV) spectra simultaneously. We applied the same  criteria for the background components and the ICM definition as explained above for the western regions. The best-fit parameters for East, North and South  are summarized in Table \ref{tab:tab8}, \ref{tab:tab7} and \ref{tab:tab9}, respectively. In general we obtained a good fit with C-stat/d.o.f. < 1.1 for E and S, and  C-stat/d.o.f. < 1.25 for N. Fig. \ref{fig:Fig8} shows the radial temperature profile for the three directions. In general, the statistical errors in our data are larger than or of the same order of magnitude as the systematic errors.

 The radial temperature profile in the E direction shows a slight increase of the temperature at the center followed by a temperature gradient until $\sim$1 keV in the cluster outer region. Motivated by a surface brightness discontinuity found at $r$~$\sim$8\arcmin\, explained in Sec. \ref{ssec:sb}, we then analyzed  a pre-shock (green annular) and post-shock (red annular) region (see Fig. \ref{fig:Fig9}). The best-fit parameters show a decrease from $kT_{\rm{post}}$ =  4.71 keV to $kT_{\rm{pre}}$ =  3.31 keV (Table \ref{tab:tab6}). The spectrum of the post-shock region is shown in Fig. \ref{fig:Fig14}. These ICM temperatures are consistent with the ones estimated in the eastern annular regions at $r$~=~6--9\arcmin\ and $r$~=~9--12\arcmin.

 For the N direction, we divided the annular regions in two (N$_{\rm{a}}$: 350-22.5\degree\  and N$_{\rm{b}}$: 22.5-55\degree) sections to investigate possible azimuthal differences. After the spectral analysis we could not find any significant difference between both sides. Therefore, we have   analyzed the full annular regions as shown in Fig.  \ref{fig:Fig7}. The temperature increases up to $\sim$5 keV at $r$~=~7.5\arcmin\ and decreases till $\sim$1 keV in the cluster outskirts. We investigated pre and post-shock regions at $r$~=~8\arcmin, similar to the case of the East direction to analyze the extent of the eastern shock, and there are no signs of a temperature drop. 

In the S direction, we obtain a radial temperature profile with a central temperature of $\sim$4 keV and a smooth decrease down to the cluster peripheries. It follows the predicted trend for relaxed galaxy clusters as described by \cite{Reiprich2013}. We discuss this behaviour in more detail in Sect. \ref{ICMstructure}. 
 
\section{Discussion of spectral analysis}

\subsection{ICM temperature profile} \label{ICMstructure}

The main growing mechanism of galaxy clusters includes the accretion and merging  of the surrounding galaxy groups and subclusters. These processes are highly energetic and turbulent, being able to modify completely the temperature structure of the ICM \citep{Markevitch2007}. Therefore, this temperature structure contains the signatures of how the cluster has evolved, providing relevant information on the growth and heating history.

 \begin{figure}[b!]
\includegraphics[width=0.5\textwidth,center]{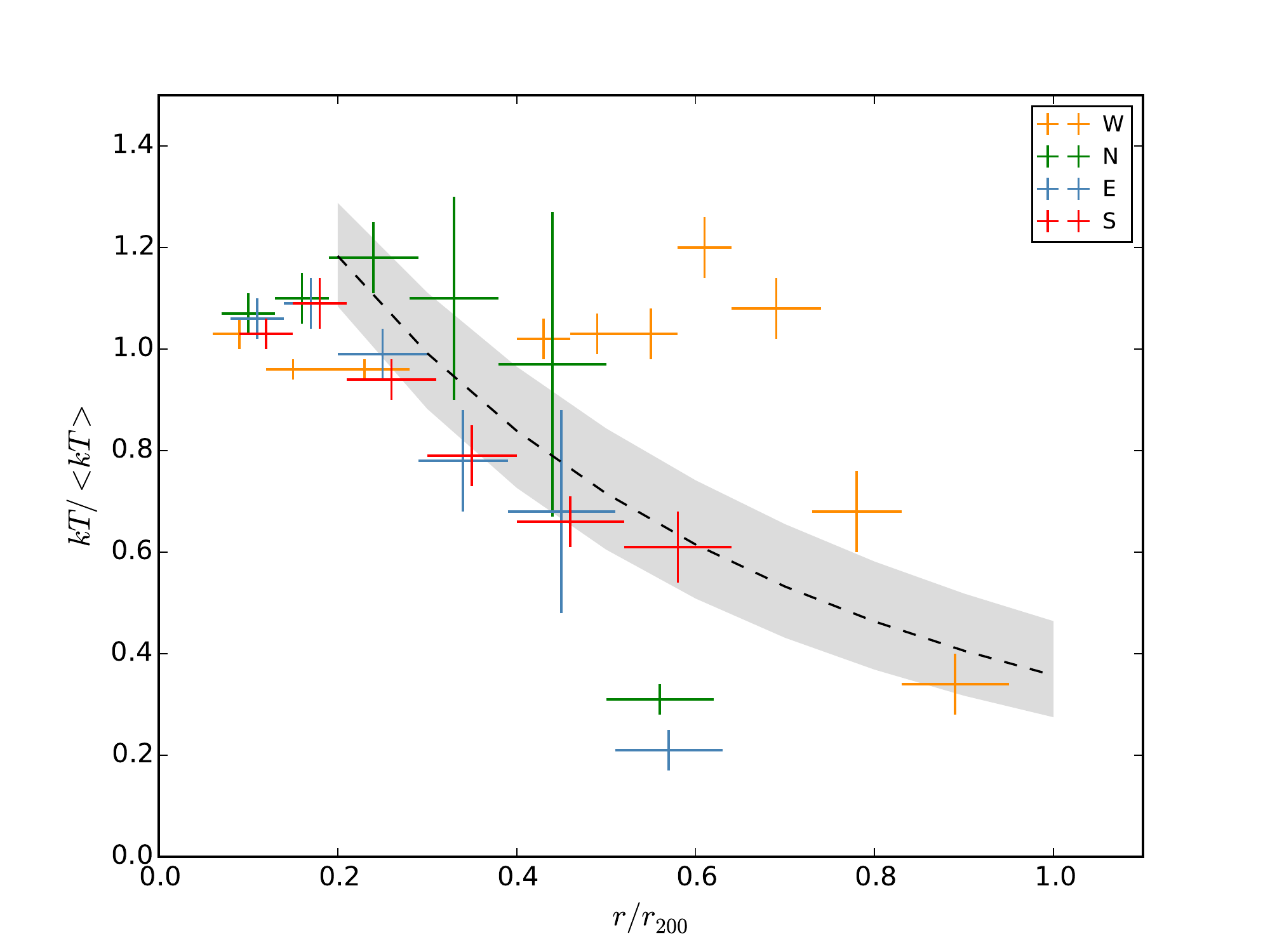}
\caption{Normalized radial temperature profile of A3376 compared with relaxed clusters. The dashed line represents the \cite{Burns2010} universal profile and the gray area shows its standard deviation. The orange, green, blue and red crosses are the scaled data for the West, North, East and South direction, respectively. We have shifted these scaled data of the different directions in $r/r_{200}$ for clarity purpose.}
\label{fig:Fig12}
\end{figure}

 There are fewer studies about merging galaxy clusters and the impact of the above events to the ICM temperature structure then about relaxed clusters.  A recent compilation of \textit{Suzaku} observations shows the temperature profile up to cluster outskirts  \citep{Reiprich2013}. That review shows that relaxed clusters have a similar behaviour near $r_{200}$ and that the temperature can smoothly drop by a factor of ~3 at the periphery. Moreover, these \textit{Suzaku} data are consistent with the ICM temperature model of relaxed galaxy clusters proposed by \cite{Burns2010}.

 \cite{Burns2010} obtained this 'universal' profile model based on N-body plus hydrodynamic simulations for relaxed clusters. The scaled temperature profile as a function of normalized radius is given by:
\begin{equation}
\label{eqn:eq7}
\frac{T}{T_{\rm{avg}}} =A\bigg[1+B\left(\frac{r}{r_{200}}\right)\bigg]^\beta,
\end{equation}
where the best-fit parameters are $A$~=~1.74~$\pm$~0.03, $B$~=~0.64~$\pm$~0.10 and $\beta$~=~3.2~$\pm$~0.4. $T_{\rm{avg}}$ is the average X-ray weighted temperature between 0.2--1.0~$r_{200}$.
 
In Fig. \ref{fig:Fig12} we compare Burns' radial profile with the radial temperature profile of A3376 normalized with $\mean{kT}$ = 4.2 keV ($\mean{kT_x}$ up to 0.3$r_{200}$, \citealt{Reiprich2009}) and $r_{200}$ $\sim$ 1.76 Mpc derived from \cite{Henry2009}. The western direction shows an enhancement of the temperature compared with the relaxed profile and a sharp drop close to $\sim$0.7~$r_{200}$. This is a  hint for the presence of a shock front and shock heating of the ICM at these radii. A similar behaviour but less pronounced is found for the North and East, being the temperature excess higher for the North. In both cases, the temperature shows a decrease around $\sim$0.3~$r_{200}$, where the eastern radio relic is located, with a steeper  temperature profile than relaxed clusters. In the  North, the large statistical errors and the weakness of the signal limit the detection of a temperature jump, and therefore, the possible evidence for a shock front. Deeper observations are needed to constrain  this temperature structure in more detail. On the other hand, the South direction seems to follow the profile of relaxed clusters. This is expected because there is no  radio emission in this direction.

 \subsection{X-ray Surface Brightness profiles} \label{ssec:sb}

\begin{figure}[ht!]
\includegraphics[width=0.5\textwidth,center]{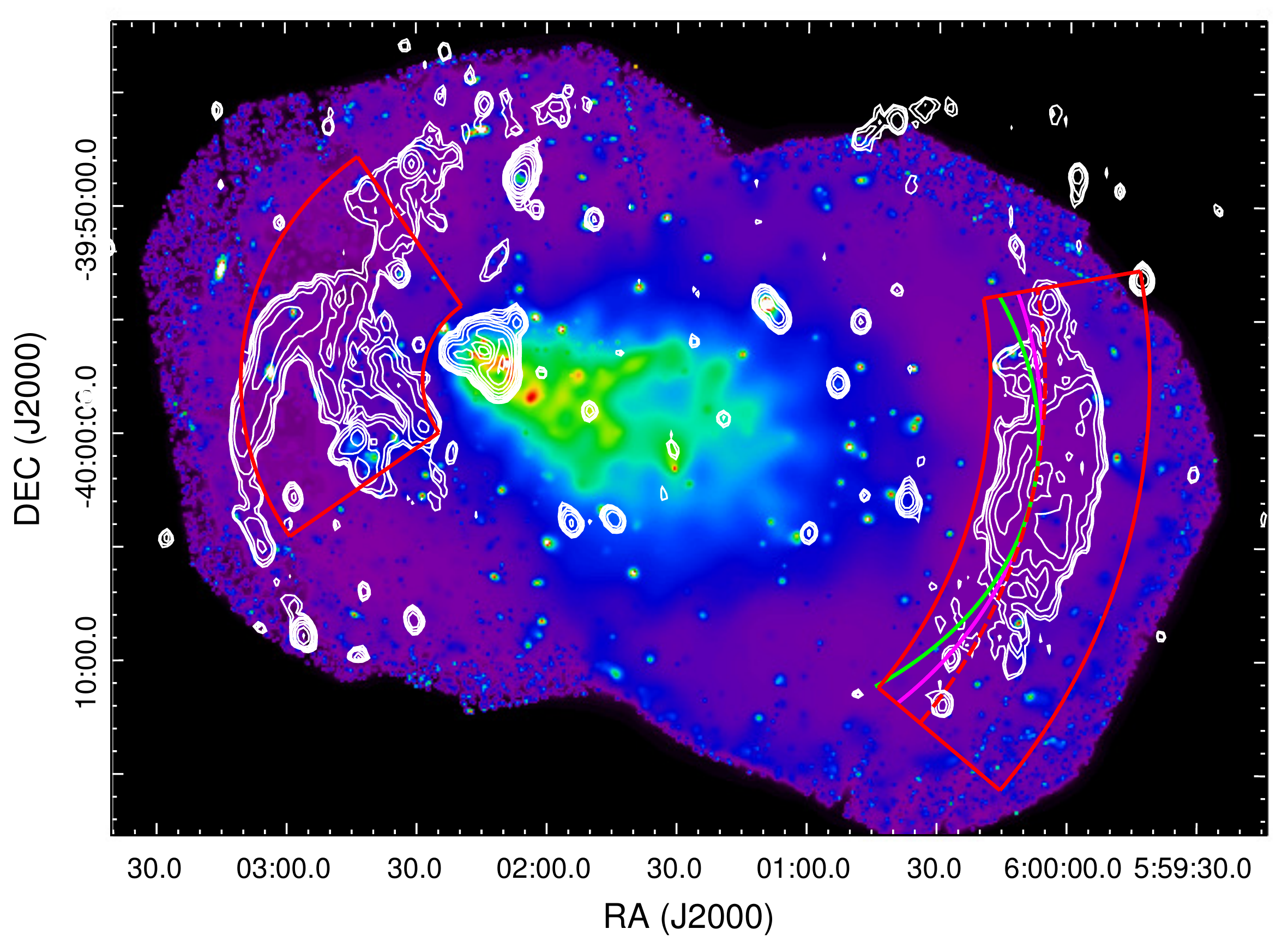}
\caption{\textit{XMM-Newton} image in the 0.5--10 keV band of A3376. The white contours correspond to VLA radio observations. The red sectors are used to extract the X-ray SB profile in the E and W directions.  The dashed red line represents the circular ($\varepsilon$~=~0) shaped sector used for the SB profiles. The magenta and green lines are the elliptical shaped sectors with $\varepsilon$~=~0.60 and $\varepsilon$~=~0.73, respectively.}
\label{fig:Fig21}
\end{figure}
 In order to confirm the evidence of bow shocks at the western and eastern regions, we analyzed the radial X-ray surface brightness profile from the center of A3376 using the \textit{XMM-Newton} observations. The SB was determined in the 0.3-2.0 keV band excluding the Al line at $\sim$1.5 keV. Point sources with a flux higher than $S_{\rm{c}}$ = 10$^{-17}$ W m$^{-2}$ were removed (for more details see Sect. \ref{bkg}) and the X-ray image was corrected for exposure and background. The sky background components derived from Q0551-3637 (see Table \ref{tab:tab2}) have been adopted together with the instrumental background for the background correction.  The surface brightness profile was extracted in circular pie shaped sectors (see sectors in Fig. \ref{fig:Fig21}) and fitted with PROFFIT v1.4\footnote{http://www.isdc.unige.ch/~deckert/newsite/Proffit.html} \citep{Eckert2011}. We adopted a broken power-law density profile to describe the SB profile. Therefore, assuming spherical symmetry, the density distribution is given by:
 \begin{eqnarray}
 \begin{cases}
n_{2}(r) = n_0\left(\frac{r}{r_{\rm{sh}}}\right)^{\alpha_2} &  r\leq r_{\rm{sh}} \\ \\
n_{1}(r) = \frac{1}{C}n_0\left(\frac{r}{r_{\rm{sh}}}\right)^{\alpha_1} & r> r_{\rm{sh}} 
\end{cases}
\end{eqnarray}
where $n_0$ is the model density normalization, $\alpha_1$ and $\alpha_2$ are the power-law indices, $r$ is the radius from the centre of the sector and $r_{\rm{sh}}$ is the shock putative distance. At the location of the SB discontinuity $n_{2}$, the post-shock (downstream) density is higher by a factor of $C=n_{2}/n_{1}$ compared with $n_{1}$, pre-shock (upstream) density. This factor $C$ is known as the compression factor.    In the X-ray SB fitting, we left all these model parameters free to vary.
 
\begin{figure}[ht!]
\includegraphics[width=0.5\textwidth,center]{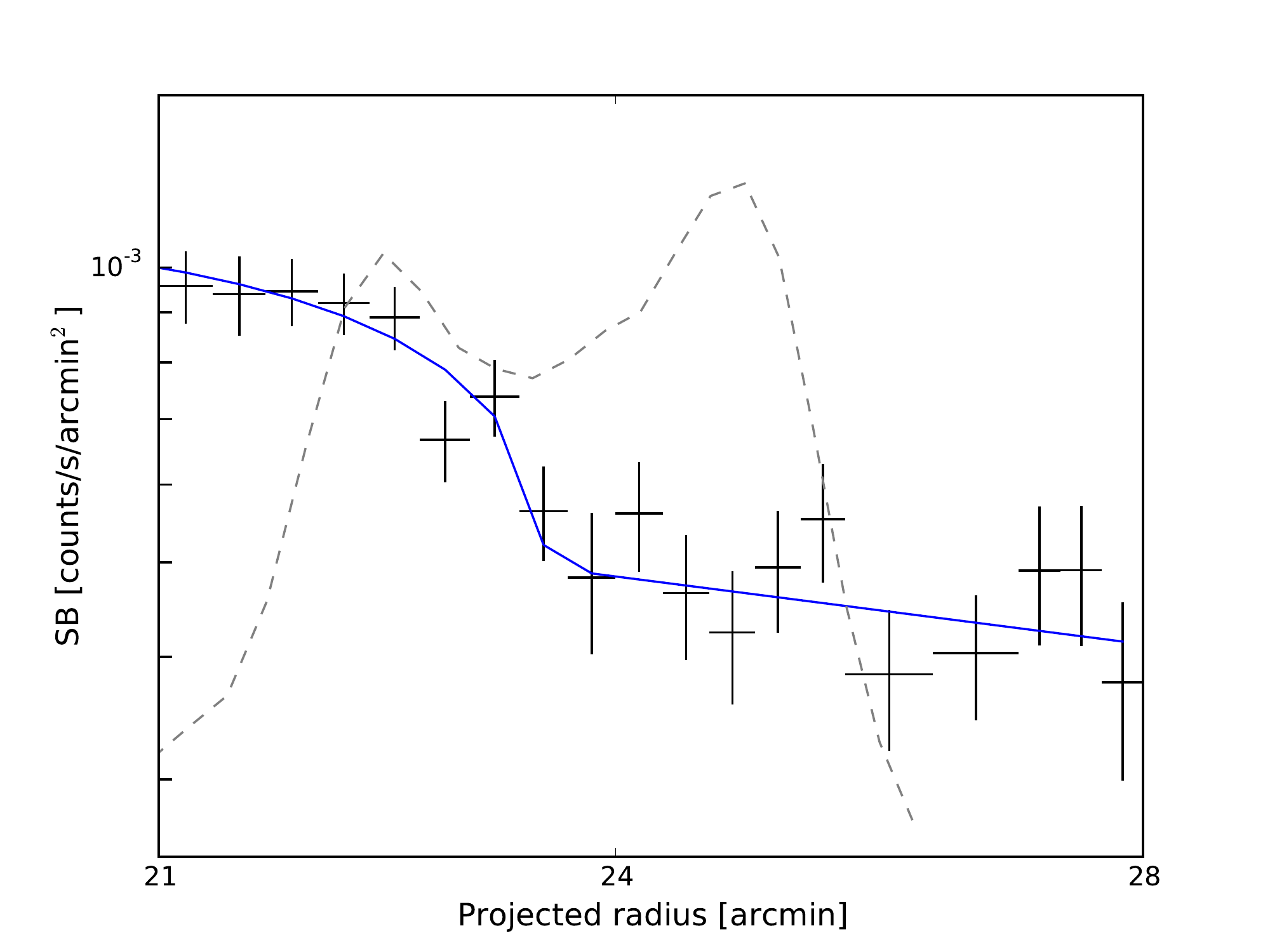}
\caption{Radial X-ray surface brightness profile across the western radio relic in the 0.3--2.0 keV band using \textit{XMM-Newton} observations. The profile is corrected for vignetting and background level, and point sources have been removed. The solid blue curve is the best-fit model and the gray dashed line is the VLA scaled radio emission. The data were rebinned to reach a minimum signal-to-noise ratio  (SNR)  of  4 and C-stat/d.o.f.~$\sim$~1.2.}
\label{fig:Fig19}
\end{figure}

 The radial SB profile across the western radio relic is shown in Fig. \ref{fig:Fig19}. The SB was accumulated in circularly shaped sectors to match the outer edge of the radio relic from an angle of 230\degree\ to 280\degree\ (measured counter-clockwise) for $r$~=~21--28\arcmin. The  data  were  rebinned  to  reach  a  minimum  signal-to-noise ratio of 4. The best-fitting broken power-law model is shown as the blue line with C-stat/d.o.f.~$\sim$~1.2. It presents a break inside the radio relic at $r_{\rm{sh}}\sim$~23\arcmin (measured from the X-ray emission peak centroid) and the compression factor is  $C$ = 1.9~$\pm$~0.4. If we apply the PSF modelling of \textit{XMM-Newton} as described in Appendix C of \cite{Eckert2016a} for a more accurate estimate of the density profile, then the density jump increases to $C$ = 2.1~$\pm$~0.6. We have additionally evaluated different elliptical shaped sectors to adjust them to the relic shape. They have eccentricities of $\varepsilon$~=~0.6 and $\varepsilon$~=~0.73. The compression factors obtained in both cases are lower than for the circular sector, $C$ = 1.7~$\pm$~0.3 and $C$ = 1.5~$\pm$~0.4, respectively.  The different radii for the SB discontinuity compared with the temperature jump could be caused by the \textit{Suzaku} PSF~$\sim$~2\arcmin.

 \begin{figure}[ht!]
\includegraphics[width=0.5\textwidth,center]{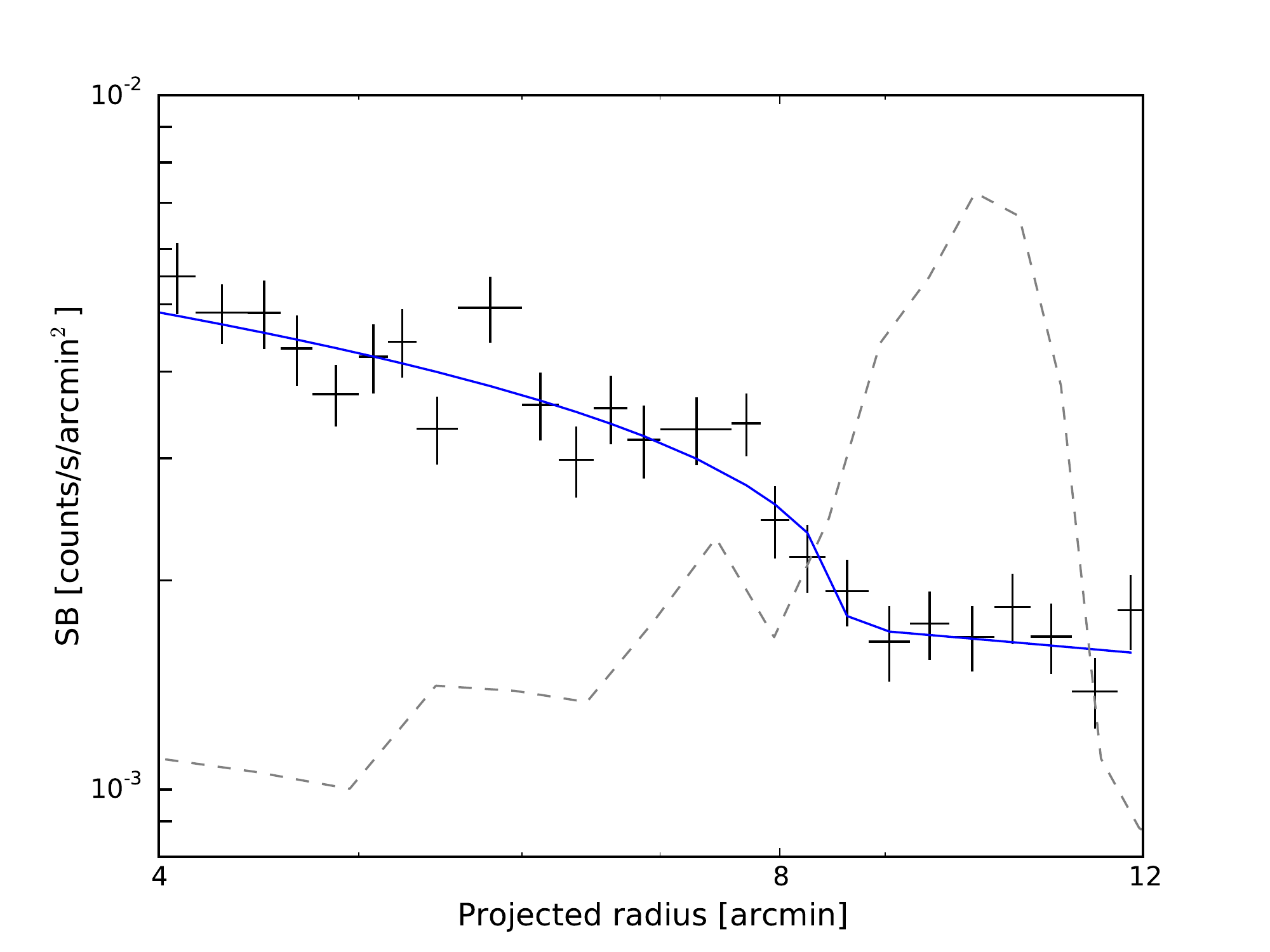}
\caption{Radial X-ray surface brightness profile across the eastern radio relic in the 0.3--2.0 keV band using \textit{XMM-Newton} observations. Same as Fig. \ref{fig:Fig19}, with a minimum  SNR of 8 and C-stat/d.o.f.~$\sim$~1.4.}
\label{fig:Fig20}
\end{figure}

 The obtained value of $C$ = 2.1~$\pm$~0.6 is consistent within the 1$\sigma$ uncertainty bounds with the value in Table \ref{tab:tab10}, $C$ = 2.9~$\pm$~0.3, although it shows a slightly lower value. This can be possibly explained because of the difficulty to model the multi-component background in the lack of any sky region where it cannot be spatially separated from the ICM. Therefore, the background modelling can play also an important role in the SB profiles located at the outskirts. Future observations with the \textit{Athena} satellite could provide an explanation to this issue.

We obtained the radial SB profile along the E direction for circular sectors between 35\degree\ to 125\degree\ with the center at the X-ray emission peak. The data were rebinned to have a minimum SNR~$\sim$~8. For radii larger than 12\arcmin\ the SB emission is low and is background dominated. For this reason, we selected the range 4--12\arcmin\ for the fitting. The best-fitting broken power-law model is shown in Fig. \ref{fig:Fig20} as a blue line with C-stat/d.o.f.~$\sim$~1.4.  The SB profile contains an edge at $r\sim$~8\arcmin\ and the compression factor is $C$ = 1.9~$\pm$~0.5, after applying the same PSF modelling as for the western SB profile. Because the radial profile of the radio relic azimutally averages various features (see VLA radio contours in Fig. \ref{fig:Fig9} and \ref{fig:Fig21}), this edge appears to be located ahead of a secondary peak in the radio relic profile. It seems to be associated to the 'notch' described by \cite{Kale2012}.

\subsection{Shock jump conditions and properties}  \label{ssec:shockprop}

The density (surface brightness) and temperature discontinuities found along the West and East direction form evidence for a shock front co-spatial with the western relic (Fig. \ref{fig:Fig19} and Table \ref{tab:tab5}) and the 'notch' radio structure in the East (Fig. \ref{fig:Fig20} and Table  \ref{tab:tab6}), respectively. Here, we calculate the shock properties at the West and East directions based on our \textit{Suzaku} observations. The Mach number (${\cal M}$) and compression factor ($C$) can  be assessed from the Rankine-Hugoniot jump condition \citep{Landau1959} assuming that all of the dissipated shock energy is thermalized and the ratio of specific heats (the adiabatic index) is $\gamma$ = 5/3:
\begin{equation}
\label{eqn:eq2}
\frac{T_2}{T_1} = \frac{5{\cal M}^4 + 14{\cal M}^2 - 3}{16{\cal M}^2} ,
\end{equation}
\begin{equation}
\label{eqn:eq3}
C = \frac{n_{2}}{n_{1}} = \frac{4{\cal M}^2}{{\cal M}^2 + 3},
\end{equation}
where $n$ is the electron density, and the indices 2 and 1 corresponds to post-shock and pre-shock regions, respectively. Table \ref{tab:tab10} shows the Mach numbers  and compression factors derived from the observed temperature jumps. The Mach numbers are in good agreement with the simulations of \cite{Machado2013} and ${\cal M}_{\rm{W}}$ is consistent with previous \textit{Suzaku} results obtained by \cite{Akamatsu2012b} (${\cal M}_{\rm{W}}$ = 3.0~$\pm~0.5$).  The Mach numbers estimated from the surface brightness discontinuities described in the previous sections are   ${\cal M}_{\rm{W_{SB}}}$~$\sim$~1.3--2.5 and ${\cal M}_{\rm{E_{SB}}}$~$\sim$~1.7~$\pm$~0.4. Both values are consistent with the value present in Table \ref{tab:tab10} within 1$\sigma$ error. The presence of shocks with ${\cal M}$~$\gtrsim3.0$ is uncommon in galaxy clusters, only four other clusters have been found with such a high Mach number ('El Gordo', \citealt{Botteon2016b}; A665, \citealt{Dasadia2016a}; CIZA J224.8+5301, \citealt{Akamatsu2015}; 'Bullet', \citealt{Shimwell2015}).

\begin{table*}[ht!]
 \begin{center}
  \caption{Shock properties at the western and  eastern radio relics.}
 \label{tab:tab10}
 \begin{threeparttable}
\begin{tabular}{cccccccc}
\hline
\hline
 \noalign{\smallskip}
&$T_2$ &  $T_1$&Mach No.&$v_{\rm{shock}}$&Compression&Power-law slope&Spectrum index\\
&(keV)&(keV)&${\cal M}$\tnote{a}&(km s$^{-1}$)\tnote{b}&$C$\tnote{c}&$p$\tnote{d}&$\alpha$\tnote{e}\\
 \noalign{\smallskip}
\hline
\noalign{\smallskip}
 W&4.22~$\pm~0.26$&1.27~$\pm~0.29$&2.8~$\pm~0.4$&1630~$\pm~220$&2.9~$\pm~0.3$&2.58~$\pm~0.22$&--0.79~$\pm~0.11$ \\
  \noalign{\smallskip}
  E&4.71~$\pm~0.42$&3.3~1$\pm~0.44$&1.5~$\pm~0.1$&1450~$\pm~150$&1.8~$\pm~0.1$&4.92~$\pm~0.85$&--1.96~$\pm~0.43$ \\
 \noalign{\smallskip}
\hline
\end{tabular}
\tiny
\begin{tablenotes}
    \item[a] ${\cal M}$ is obtained from eq. (\ref{eqn:eq2}).
    \item[b]  $v_{\rm{shock}}$~=~${\cal M}$~$\cdot$~$c_{\rm{s}}$, $c_{\rm{s}}$ = $\sqrt{\gamma kT_1/\mu m_p}$
    \item[c] $C$ from eq. (\ref{eqn:eq3}).
    \item[d] $p~=~(C~+~2)/(C~-~1)$
    \item[e] $\alpha~=~-(p-1)/2$
\end{tablenotes}
 \end{threeparttable}
 \end{center}
\end{table*}

The sound speed at the pre-shock regions is $c_{\rm{s,W}}$~$\sim$~570 km s$^{-1}$ and  $c_{\rm{s,E}}$~$\sim$~940 km s$^{-1}$, assuming $c_{\rm{s}}$ = $\sqrt{\gamma kT_1/\mu m_p}$ where $\mu$ = 0.6. The shock propagation speed $v_{\rm{shock}}$~=~${\cal M}$~$\cdot$~$c_{\rm{s}}$ for the western and eastern direction is $v_{\rm{shock},W}$~$\sim$~1630~$\pm~220$ km s$^{-1}$ and $v_{\rm{shock},E}$~$\sim$~1450~$\pm~150$ km s$^{-1}$, respectively. These velocities are consistent with previous work for A3367 W \citep{Akamatsu2012b}. However, the shock velocities are smaller than in other galaxy clusters with ${\cal M}$~$\sim$~2--3 as the Bullet cluster (4500 km s$^{-1}$, \citealt{Markevitch2002}), CIZA J2242.8+530 (2300 km s$^{-1}$, \citealt{Akamatsu2017}) and A520 (2300 km s$^{-1}$, \citealt{Markevitch2005}).

\begin{figure}[t!]
\includegraphics[width=0.5\textwidth,center]{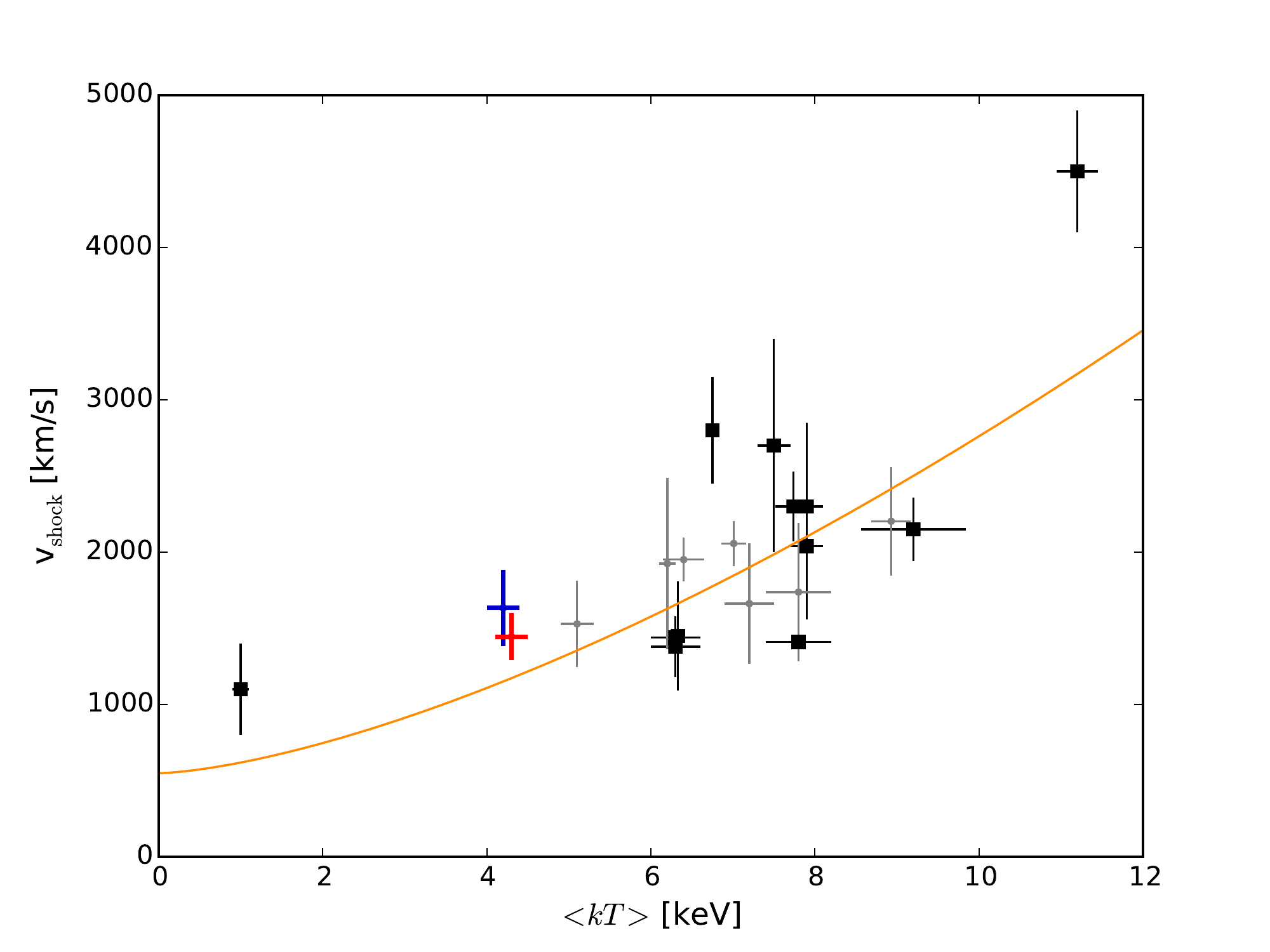}
\caption{$v_{\rm{shock}}$ correlation with $\mean{kT}$. The orange line is the best-fit for $v_{\rm{shock}}$ = $A*\mean{kT}^{(3/2)} + B$. Black crosses represent  data taken directly from the literature \citep{Akamatsu2013a,Akamatsu2015,Akamatsu2017,Dasadia2016a,Eckert2016b,Markevitch2002,Markevitch2005,Russell2012,Russell2014,Sarazin2016}. Gray points are the calculated data from ${\cal M}$ and $kT_{1}$ \citep{Botteon2016a,Bourdin2013,Macario2011,Ogrean2013,Owers2014,Sarazin2016,Trasatti2015} as explained in the text. Blue and red crosses correspond to the W and E shocks of A3376 as obtained in this work, respectively.}
\label{fig:Fig15}\label{shockprop}
\end{figure}

We have compiled the shock velocities, $v_{\rm{shock}}$, for several merging galaxy clusters as shown in Fig. \ref{fig:Fig15}. Black points represent data taken directly from the literature and the gray points are calculated from ${\cal M}$ and $kT_{1}$. Blue and red crosses are the western and eastern shock velocities, respectively. Fig. \ref{fig:Fig15} shows shock velocities as a function of average temperature of the system. To investigate the origin of driving force of the shock structure, we have examined  a prediction from self-similar relationship: $v_{\rm{shock}}$ = $A*\mean{kT}^{(3/2)} + B$, where $A$~=~70~$\pm$~16 and $B$~=550~$\pm$~270, see orange line. As we expected, most of samples can be explained with this formula, which means that the
main driving force of the merging activity is the gravitational potential of the system  and no additional physics. Further sample of shocks and proper gravitational mass estimates will enable us to extend this type of examination.

\subsection{Mach number from X-ray and radio observations}

After the first confirmation by \cite{Finoguenov2010} of a clear correlation between X-ray shock fronts and radio relics in A3667, many other observations have revealed a possible relation between them \citep{Macario2011,Mazzota2011,Akamatsu2013a}. As explained in the introduction, the X-ray shock fronts may accelerate charged particles up to relativistic energies via diffusive shock acceleration \citep[DSA,][]{Drury1983,Blandford1987}, which in the presence of a magnetic field can generate synchrotron emission. The acceleration efficiency of this mechanism is low for shocks with ${\cal M}$ < 10 and might not be sufficient to produce the observed radio spectral index  \citep{Kang2012,Pinzke2013}. Therefore, alternative scenarios have been proposed like the presence of a fossil population of non-thermal electrons and re-acceleration of these electrons by shocks \citep{Bonafede2014,VanWeeren2017}  or the electron re-acceleration by turbulence \citep{Fujita2015,Kang2017}.

DSA is based on first order Fermi acceleration, considering that there is a stationary and continuous injection, which accelerates relativistic electrons following a power-law spectrum $n(E)dE$ $\sim$ $E^{-p}dE$ with
$p~=~(C~+~2)/(C~-~1)$, $\alpha~=~-(p-1)/2$ 
where $p$ is the power-law index and $\alpha$ is the radio spectral index for $S_v$~$\varpropto$~$v^{\alpha}$. The radio Mach number can be calculated from the injection spectral index as:
\begin{equation}
\label{eqn:eq4}
{\cal M}_R^2 = \frac{2\alpha-3}{2\alpha + 1}. 
\end{equation}

For A3376 two radio observations exist in addition to the Very Large Array (VLA) observations by \cite{Bagchi2006}: Giant Metrewave Radio Telescope, GMRT, 150 and 325 MHz: \citealt{Kale2012} and Murchison Widefield Array, MWA, 80-215 MHz: \citealt{George2015}.
\cite{Kale2012} describe in detail the morphology of the E and W relics. They consider the DSA mechanism assuming that the synchrotron and inverse Compton losses have not affected the spectrum. Therefore, they use for the injected spectral index calculation the flattest spectrum in the outer edge of the spectral index map for 325--1400 MHz. Their results are $\alpha_{\rm{E}}$ = --0.70~$\pm$~0.15, which implies ${\cal M}_{\rm{E}}$ = 3.31 $\pm$ 0.29; and $\alpha_{\rm{W}}$ = --1.0~$\pm$~0.02 with ${\cal M}_{\rm{W}}$ = 2.23 $\pm$ 0.40. In this case, there are slight differences between the Mach number assessed from the X-ray and radio observations. \cite{George2015} have derived the integrated spectral indices of the eastern and western relic using the GMRT, MWA and VLA observations (see their Fig. 3) obtaining $\alpha_{\rm{E}}$ = --1.37~$\pm$~0.08 and $\alpha_{\rm{W}}$ = --1.17~$\pm$~0.06, which give the Mach numbers, assuming $\alpha_{\rm{int}}$~=~$\alpha_{\rm{inj}}$ -- 0.5, ${\cal M}_{\rm{E}}$ = 2.53 $\pm$ 0.23 and ${\cal M}_{\rm{W}}$ = 3.57 $\pm$ 0.58, respectively. However, due to the low angular resolution of MWA, they can only resolve the integrated spectral index and not the injected one. For this reason, we have decided not to use the \cite{George2015} results for this study. 
   
\begin{figure}[b!]
\includegraphics[width=0.5\textwidth,center]{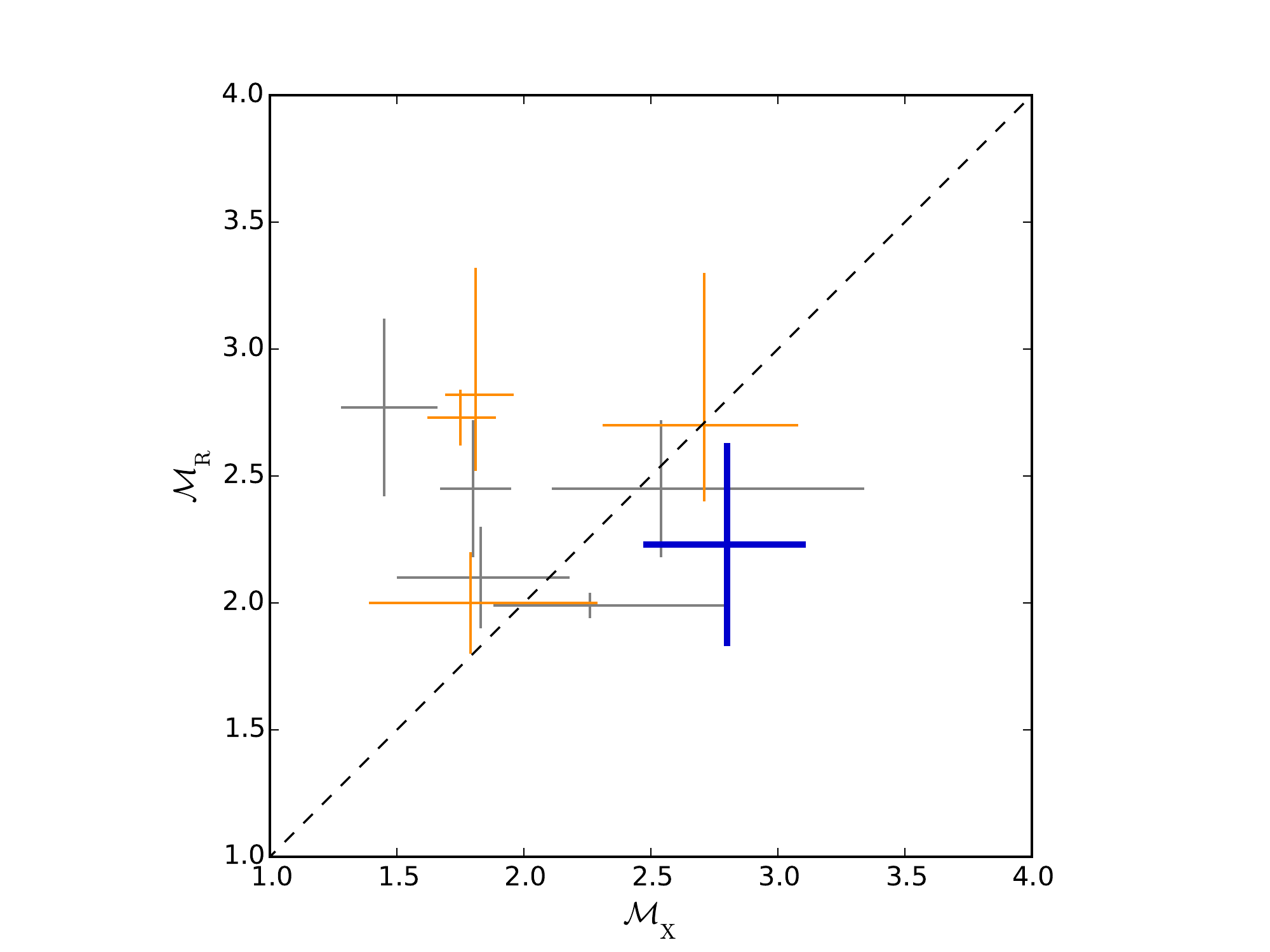}
\caption{Comparison between Mach number derived from radio observations (${\cal M}_{\rm{R}}$) based on the radio spectral index and the X-ray observations  (${\cal M}_{\rm{X}}$). The results for A3376 W is represented by the blue cross. The gray crosses show data for others clusters (see Table \ref{tab:tab11}), orange crosses are recent radio observations done with LOFAR references. The black dashed line is the linear correlation used as reference.}
\label{fig:Fig10}
\end{figure}

The differences between ${\cal M}_{\rm{R}}$ and ${\cal M}_{\rm{X}}$ are not only present in this galaxy cluster. Fig. \ref{fig:Fig10} includes several systems with similar behaviour as A3376 (see Table \ref{tab:tab11}), which is shown by a blue cross for A3376 W. A3376 E is not included because the spectral index reported by the radio observations refers to the elongated radio relic and not to the 'notch' structure. The orange crosses represent radio data obtained with LOFAR. ${\cal M}_{\rm{X}}$ of these clusters is derived from the temperature jump and ${\cal M}_{\rm{R}}$ is calculated using eq. (\ref{eqn:eq4}) with the injection spectral index $\alpha_{\rm{inj}}$. In addition, some other clusters as A521 \citep{Bourdin2013,Giacintucci2008}, A2034 \citep{Owers2014,Shimwell2016}  or 'El Gordo' \citep{Botteon2016b, Lindner2014};  present evidence for a shock front based on the surface brightness jump.  In order to remain consistent regarding the method used to derive ${\cal M}$ (in our case, the temperature jump), we do not include these measurements in Fig. \ref{fig:Fig10}. We note that Mach numbers based on the SB jump seem to be smaller than those based on temperature jump (for more details, see \cite{Sarazin2016}).
 
\begin{table}[b!]
 \begin{center}
  \caption{The cluster sample for our ${\cal M}_{\rm{X}}$ and ${\cal M}_{\rm{R}}$ comparison.}
  \label{tab:tab11}
  \begin{threeparttable}
 \begin{tabular}{ccc}
\hline 
\hline
\noalign{\smallskip}
 Cluster &${\cal M}_{\rm{X}}$\tnote{$\dagger$}&${\cal M}_{\rm{R}}$\\ 
 \noalign{\smallskip}
 \hline
 \noalign{\smallskip}
  A2256\tnote{*}&1.8~$\pm$~0.1 (1)&2.7~$\pm$~0.1 (2)\\
 \noalign{\smallskip}
  A2255 NE&1.5~$\pm$~0.2 (3)&2.8~$\pm$~0.4 (4)\\
 \noalign{\smallskip}
 A115& 1.8~$\pm$~0.3 (5)&2.1~$\pm$~0.2 (5)\\
 \noalign{\smallskip}
   Sausage N\tnote{*}&  2.7~$\pm$~0.4 (6) &2.7~$\pm$~0.5 (8)\\
 \noalign{\smallskip}
    Sausage S\tnote{*}& 1.8~$\pm$~0.5 (7) &2.0~$\pm$~0.2 (8)\\
 \noalign{\smallskip}
    A3667 NW& 2.5$^{+0.8}_{-0.4}$ (9) &2.5~$\pm$~0.3 (10)\\
 \noalign{\smallskip}
  A3667 SE& 1.8~$\pm$~0.1 (11)&2.5~$\pm$~0.3 (10)\\
 \noalign{\smallskip}
    Coma& 2.3~$\pm$~0.5 (12) &1.99~$\pm$~0.05 (13)\\
 \noalign{\smallskip}
    Toothbrush\tnote{*}& 1.8~$\pm$~0.2 (14) &2.8~$\pm$~0.4 (15)\\
 \noalign{\smallskip}
    A3376 W& 2.8~$\pm$~0.4 &2.2~$\pm$~0.4 (16)\\
 \noalign{\smallskip}
 
 \hline
 \end{tabular}
 \tiny
\begin{tablenotes}
    \item[*]Radio observations with LOFAR.
    \item[$\dagger$]${\cal M}_{\rm{X}}$ has been calculated as described by \cite{Sarazin2016}.\\
    
\end{tablenotes}
 \end{threeparttable}
  
  \begin{minipage}{7.5cm}%
  \tiny
  References: (1) \cite{Trasatti2015}; (2)  \cite{VanWeeren2012};
  (3) \cite{Akamatsu2017}; (4) \cite{Pizzo2009}; (5) \cite{Botteon2016a}; (6) \cite{Akamatsu2015};  (7) \cite{Storm2017};  (8) \cite{Hoang2017}; (9) \cite{Sarazin2016}; (10) \cite{Hindson2014}; (11) \cite{Akamatsu2013a}; (12) \cite{Akamatsu2013b}; (13) \cite{Thierbach2003};  (14) \cite{Itahana2015}; (15) \cite{VanWeeren2016}; (16) \cite{Kale2012}.
  \end{minipage}

  \end{center}
 \end{table}

\cite{Akamatsu2017} discuss several possible reasons for the Mach numbers discrepancies in X-rays and radio. In this study we focus on the radio ageing effect \citep{Pacholczyk1970, Miniati2002,Stroe2014}, which can lead to lower radio ${\cal M}$ compared to X-rays. The electron ageing effect takes places when the relativistic electrons lose their energy via radiative cooling or inverse-Compton scattering after the shock passage in approximately less than $\sim$10$^7$--10$^8$ years, which is shorter than the shock life time. As a consequence, the radio spectrum becomes steeper and the integrated spectral index decreases from  $\alpha_{\rm{int}}$ to $\alpha_{\rm{inj}}$ -- 0.5  in DSA model \citep{Pacholczyk1970, Miniati2002}. Only measurements with high angular resolution and low-frequency are able to measure $\alpha_{\rm{inj}}$ directly. As a consequence, the calculation of ${\cal M}_{\rm{R}}$ might be underestimated. Moreover, it is thought that the simple relationship of $\alpha_{\rm{int}}$ = $\alpha_{\rm{inj}}$ -- 0.5 most likely does not hold for the relics case \citep{Kang2015,kang2016}. In the case of A3376, \cite{Kale2012} showed that the spectral distribution steepens from the outer to the inner edge in the frequency range 325--1400 MHz. However, to avoid the ageing effect they consider only the flattest spectral index as $\alpha_{\rm{inj}}$ at the outer edge.  

We calculate the distance free of ageing effect as $d$~=~$t_{\rm{loss}}$~$\times$~$v_{\rm{gas}}$, where   $t_{\rm{loss}}$ is the cooling time for relativistic electrons and $v_{\rm{gas}}=v_{\rm{shock}}/C$ is the gas velocity. We estimate $t_{\rm{loss}}\sim$~10--5~$\times$~10$^7$ yr from Eq. (14) of \cite{Kang2012}:
\begin{equation}
\label{eqn:eq5}
t_{loss} \approx 8.7~\times~10^8~\rm{yr}~ \Bigg(\frac{B^{1/2}}{B^2_{\rm{eff}}}\Bigg)~\Bigg(\frac{v_{\rm{obs}}}{1~\rm{GHz}}\Bigg)^{-1/2}(1+z)^{-1/2},
\end{equation}
where we assume a magnetic field of B~$\sim$~1$~\mu$G and $v_{\rm{obs}}$~=~325 MHz--1.4 GHz. The effective magnetic field is B$^2_{\rm{eff}}$~=~B$^2$+B$^2_{\rm{CBR}}$, which includes the equivalent strength of
the cosmic background radiation like B$_{\rm{CBR}}$~=~3.24~$\mu$G~(1+$z$)$^2$.  The gas velocity is $v_{\rm{gas,W}}\sim$~564 km/s for the West. Therefore, the distance free of the ageing effects is $d_{\rm{W}}$~$\sim$58--28 kpc. In the case of assuming a five times higher magnetic field, B~$\sim$~5$~\mu$G, $t_{\rm{loss}}\sim$~4~$\times$~10$^7$ yr (20 \% lower than with B~$\sim$~1$~\mu$G) and for B~$\sim$~0.2$~\mu$G $t_{\rm{loss}}\sim$~2~$\times$~10$^7$ yr (60 \%), for $v_{\rm{obs}}$~=~1.4 GHz. Therefore, the assumptions of B~$\sim$~1$~\mu$G seems the most conservative one.  This distance at 1.4 GHz, 28 kpc, is smaller than the radio beam at same frequency  (69\arcsec~$\times$~69\arcsec or 65~$\times$~65 kpc) used for the spectral index estimation \citep{Kale2012}. Thus, ${\cal M}_{\rm{R}}$ might be affected by the ageing effect.

Additionally, the X-ray SB discontinuities in the West and East (see Figs. \ref{fig:Fig19} and \ref{fig:Fig20}) do not coincide with the outermost edges of the radio relics. There is significant radio structure outside these SB discontinuities. Therefore, while the DSA mechanism associated to these shocks is assumed to be responsible for the radio emission directly behind the shocks, it cannot explain all radio emission. There might be other shocks further outwards, but these are below the detection threshold using current instruments, due to a combination of low spatial resolution (\textit{Suzaku}) or high background (\textit{XMM-Newton}), so we cannot test that hypothesis in the present study.
Together with the ageing effect discussed before, this might contribute to the difference between the Mach number derived from the X-ray and radio observations.

\subsection{Merger scenario}

The N-body/SPH simulations of A3376 by \cite{Machado2013} predict a merger scenario with two subclusters: one, eastern, which is more compact with four times more concentrated gas than the other, western subcluster. The eastern subcluster  has a high initial velocity and is able to cross through the more massive western subcluster, disrupting its core and forming the dense and bright tail. This could be the reason why only one X-ray emission peak is found, probably associated to the BCG2 of the Eastern subcluster in the central region. This scenario is confirmed by the weak lensing analysis of \cite{Monteiro-Oliveira2017}, which reveals that the mass peak concentration is in the stripped tail. 

From the shock properties, we are able to derive the dynamical age of A3376. Assuming that the western and eastern shocks have travelled from the cluster core to the radio relic location with respective constant velocity ($v_{\rm{shock,W}}$~$\sim$~1630~$\pm~220$ km s$^{-1}$  and  $c_{\rm{shock,E}}$~$\sim$~1450~$\pm~150$ km s$^{-1}$) and the distance between both shocks is $\sim$1.9 Mpc, the time required to reach the current position is $\sim$0.6 Gyr. This value is in good agreement with the previous estimates by \cite{Monteiro-Oliveira2017}, \cite{George2015} and \cite{Machado2013}.  It might indicate that A3376 is a young merger cluster which is still evolving and following the outgoing scenario as proposed by \cite{Akamatsu2012b} and \cite{Monteiro-Oliveira2017}.  

We also estimate  the inclination angle of the line-of-sight with respect to the merging axis, assuming that the galaxies of the infalling (Eastern) subcluster are moving together with the shock front. The brightest galaxy of the E subcluster has a $z$~=~0.045591~$\pm$~0.00008 \citep{Smith2004} and the peculiar velocity with respect to the entire merging cluster ($z$~=~0.0461~$\pm$~0.003, \citealt{Monteiro-Oliveira2017}) is $\sim$~154~$\pm$~94 km/s. From the relation $\theta$~=~arctan($v_{\rm{spec}}/v_{\rm{shock,E}}$), we estimate  $\theta$~$\sim$~6\degree$\pm$~4\degree. This means that the merger axis is close to the plane of sky. 

\begin{figure}[b!]
\includegraphics[width=0.40\textwidth,center]{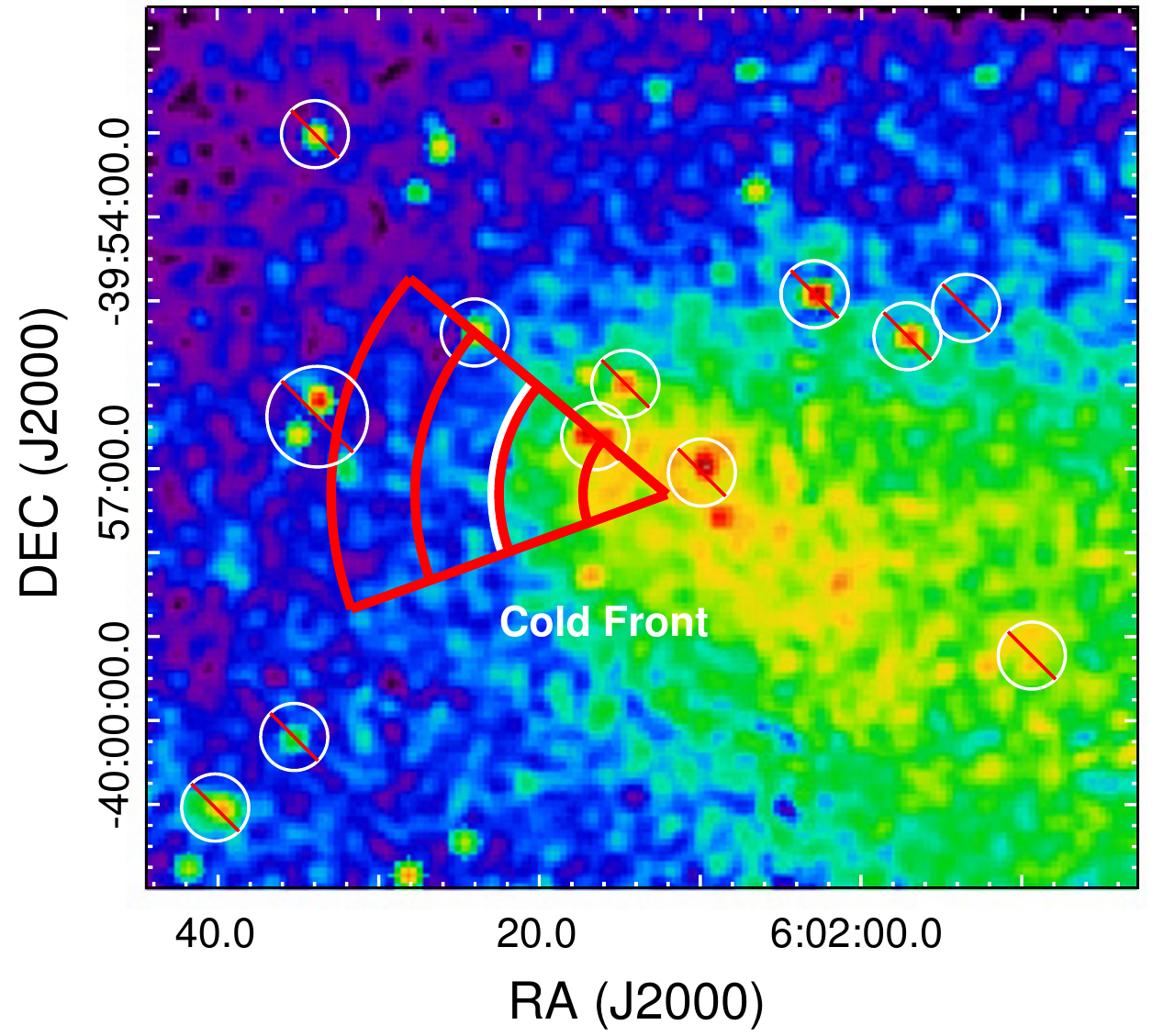}
\caption{A3376  \textit{Chandra} background and exposure corrected image in the 0.5--2.0 keV band. The red annular regions are used in the spectral analysis of the cold front and the same pie shaped annuli are used for the SB profile estimation.}
\label{fig:Fig16}
\end{figure}

\section{Cold front near the center}

\begin{figure}[t!]
 \begin{minipage}[t]{1\textwidth}
\includegraphics[width=0.5\textwidth]{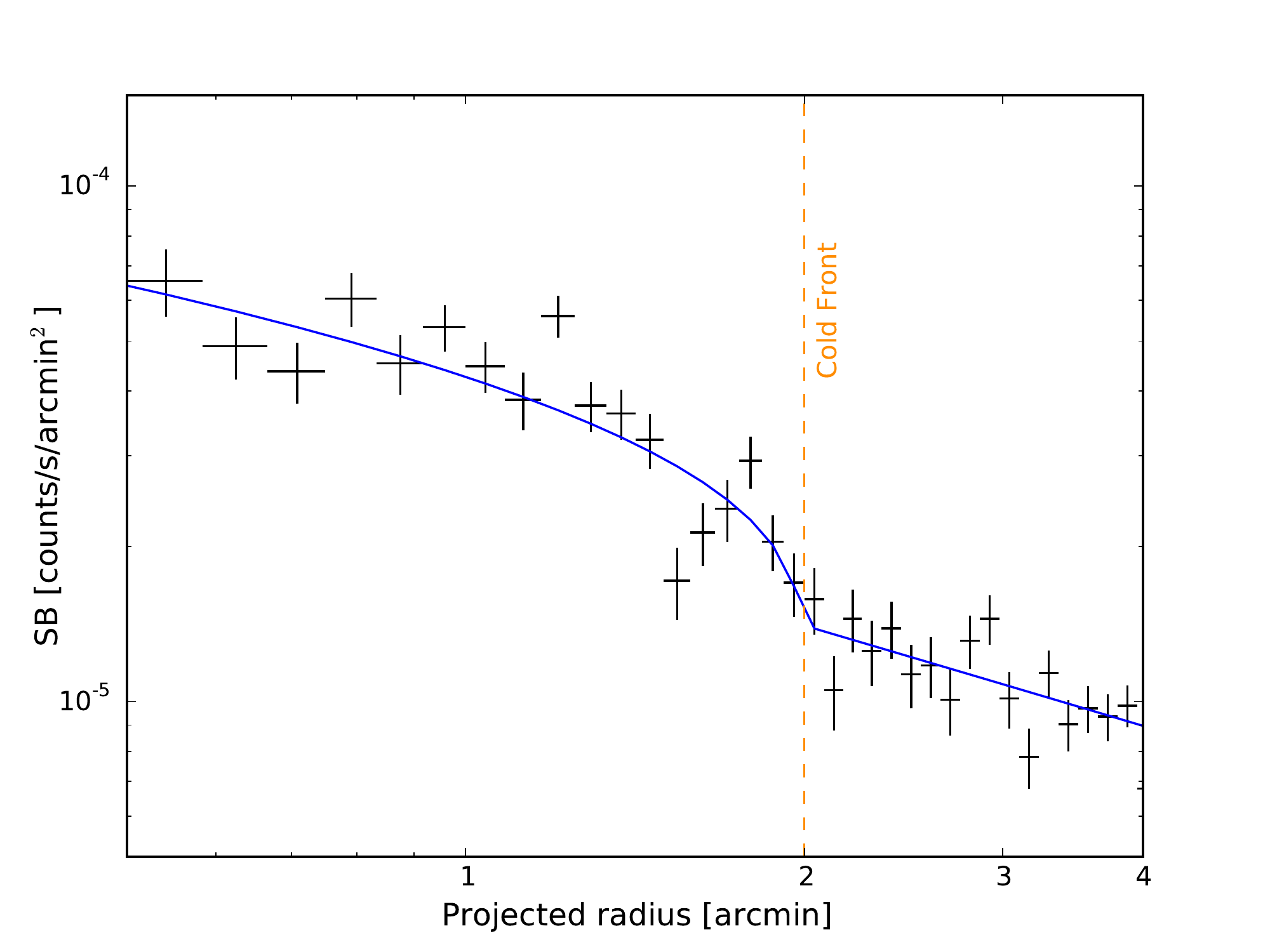}
\end{minipage}
\caption{Radial X-ray surface brightness profile across the cold front in the 0.5-2.0 keV band using the \textit{Chandra} observation. The profile is corrected for exposure and background, and point sources are removed. The solid blue curve is the best-fit model and the vertical orange line is the estimated position of the cold front. }
\label{fig:Fig17}
\hfill
\begin{minipage}[t]{1\textwidth}
\includegraphics[width=0.5\textwidth]{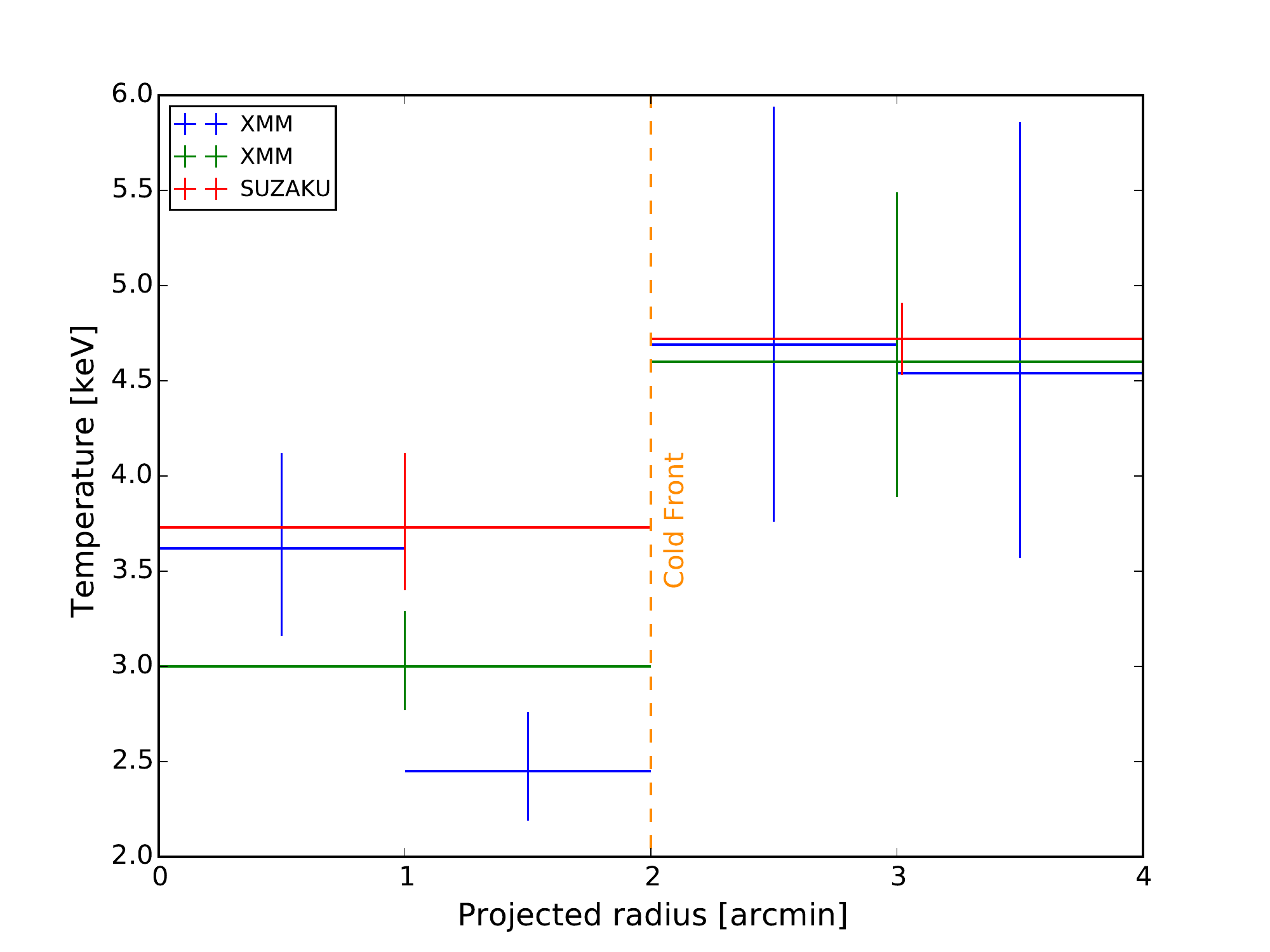}
\end{minipage}
\caption{Radial temperature  distribution for the red annular regions shown in Fig. \ref{fig:Fig16}. Blue and green crosses correspond to \textit{XMM-Newton} observations and red crosses to \textit{Suzaku}. The vertical orange line is the estimated position of the cold front. }
\setlength{\unitlength}{5cm}
\label{fig:Fig18}
  \end{figure}

The \textit{Chandra} image shown in Fig. \ref{fig:Fig16} reveals an arc-shape edge between 25--140\degree\ (measured counter-clockwise), orthogonal to the merging axis at NE direction.  For a refined spatial analysis of this feature, we extracted the X-ray surface brightness profile from the center of A3376 to the E using the \textit{Chandra} observations (see Table  \ref{tab:tab1})  in the 0.5--2.0 keV band. The circular pie annuli used to accumulate the SB profile cover the angle interval 50--120\degree\ from the centroid in $\rm{RA}=6^{\rm{h}} 02^{\rm{m}} 12\fs08$, $\rm{Dec.}=-39\degr 57\arcmin 18\farcs 53$ (same as the red annular sectors of  Fig. \ref{fig:Fig16}). The same regions are used for the spectral analysis where point sources using the criteria of $S_{\rm{c}}$ = 10$^{-17}$ W m$^{-2}$ have been excluded. Fig. \ref{fig:Fig17} shows the radial SB profile along the E direction. It shows an edge ($C$~=~1.78~$\pm$~0.15) in the SB profile at approximately 150 kpc ($\sim$3\arcmin) distance from the X-ray emission peak.

We determined the radial temperature profile using \textit{XMM-Newton} and \textit{Suzaku} observations, see Fig. \ref{fig:Fig18},  across the E direction for the four annular regions shown in Fig. \ref{fig:Fig16}.  The temperature rises from $\sim$3.0 to $\sim$4.6 keV ($T_{\rm{in}}/T_{\rm{out}}$~=~0.65~$\pm$~0.08) just outside of the SB edge location based on \textit{XMM-Newton} observations.

The radial SB and temperature profiles show that the gas is cooler and denser (brighter) inside the edge. Using the density and temperature discontinuities of $C$~=~1.78~$\pm$~0.15 and $T_{\rm{in}}/T_{\rm{out}}$~=~0.65$~\pm~$0.08, respectively; the pressure jump is $P_{\rm{in}}/P_{\rm{out}}$~=~1.16~$\pm$~0.17. Therefore, the pressure is nearly continuous across the edge, which agrees with the definition of a cold front. 

\begin{table}[t]
 \begin{center}
 \caption{Temperature distribution at cold front regions shown in Fig. \ref{fig:Fig17}.}
  \label{tab:tab12}
 \begin{tabular}{ccc}
  \hline
\hline
\noalign{\smallskip}
 &Radius (\arcmin)&\textit{kT} (keV)\\ 
 \noalign{\smallskip}
 \hline
 \noalign{\smallskip}
&0.5~$\pm~0.5$& 3.62~$\pm~0.46$ \\
 \noalign{\smallskip}
\textit{XMM-Newton} &1.5~$\pm~0.5$& 2.45~$\pm~0.29$\\
 \noalign{\smallskip}
&2.5~$\pm~0.5$& 4.69~$\pm~1.14$\\
 \noalign{\smallskip}
&3.5~$\pm~0.5$& 4.54~$\pm~1.15$\\
 \noalign{\smallskip}
 \hline
   \noalign{\smallskip}
 \textit{XMM-Newton}&1.0~$\pm~1.0$& 3.00~$\pm~0.27$\\
 \noalign{\smallskip}
 &3.0~$\pm~1.0$& 4.60~$\pm~0.80$\\
 \noalign{\smallskip}
  \hline
  \noalign{\smallskip}
\textit{Suzaku} &1.0~$\pm~1.0$& 3.73~$\pm~0.36$\\
 \noalign{\smallskip}
 &3.0~$\pm~1.0$& 4.71~$\pm~0.19$\\
 \noalign{\smallskip}
  \hline
 \end{tabular}
  \end{center}
 \end{table}
 
 \subsection{Cold front properties}

In addition to the two shock fronts described in the previous sections, we found evidence of a sharp edge close to the center associated to a cold front. It delimits the boundaries of a cool and dense gas cloud moving through a hotter  ambient gas, which is subject to ram pressure stripping \citep{Markevitch2007,Vikhlinin2001}.  In this case, Fig. \ref{fig:Fig16} shows this shape of the front perpendicular to the merger axis advancing in the NE direction.   
Using the temperatures and densities derived from the X-ray image and spectral analysis, we estimate the velocity of the cool gas cloud \citep{Landau1959,Vikhlinin2001,Dasadia2016b,Ichinohe2017}. We approximate the cool gas as a spherical body advancing into ambient gas. If the cloud moves with supersonic velocity, a shock front could form upstream of the gas cloud. Near the cold front edge the cool gas decelerates to zero velocity at the stagnation point, referred here with the index 0.  We assume as well a free stream of gas upstream of the cold front, which is undisturbed and flows free, denoted by index 1. The ratio of pressures in the free stream and the stagnation point can be expressed as a function of the gas cloud speed $v$ \citep{Landau1959}: 
\begin{equation}
  \frac{p_0}{p_1}=\begin{cases}
\Big(1+\frac{\gamma-1}{2}{\cal M}_1^2\Big)^{\gamma/(\gamma-1)} & ({\cal M}_1\leq1)\\
\\
  \Big(\frac{\gamma+1}{2}\Big)^{(\gamma+1)/(\gamma-1)}{\cal M}_1^2\Big(\gamma-\frac{\gamma-1}{2{\cal M}_1^2}\Big)^{-1/(\gamma-1)}   & ({\cal M}_1>1)
  \end{cases}
\end{equation}
where $\gamma$~=~5/3 is the adiabatic index; ${\cal M}_1$~=~$v$/$c_{\rm{s}}$  and $c_{\rm{s}}$ are the Mach number and sound velocity of the free stream, respectively. 

The gas parameter at the stagnation point cannot be measured directly, although the stagnation pressure is similar to the gas cloud inside the cold front \citep{Vikhlinin2001}. Therefore we assume $T_0$~=~3.00~$\pm$~0.27 keV as the temperature inside the cold front  and  $T_1$~=~4.60~$\pm$~0.80 keV  outside the cold front used as the free stream region \citep{Vikhlinin2001, Sarazin2016}. The density ratio of these regions is calculated from the best-fit parameter of the broken power-law model of Fig. \ref{fig:Fig17}, $n_{0}/n_1$~=~3.91~$\pm$~0.75.  Thus, the pressure ratio between the inside of the cold front and the outside free stream region is $p_0/p_1$~=~~2.6~$\pm$~0.6, which corresponds to ${\cal M}_1$~=~1.2~$\pm$~0.2 in the free stream. Using $c_{\rm{cf}}$ = $\sqrt{\gamma kT_1/\mu m_p}$ with $kT_1$~=~4.60 keV, $c_{\rm{cf}}$~=~1100~$\pm$~100 km s$^{-1}$ and  the velocity of the cool gas cloud is $v_{\rm{cf}}$~=~1300~$\pm$~250 km s$^{-1}$. This value is consistent with the velocity of the shock, $v_{\rm{cf}}$~=~1450~$\pm$~150 km s$^{-1}$, front derived in Sec. \ref{ssec:shockprop}. 

 As explained in \cite{Dasadia2016b}, for a rigid sphere the ratio of $d_{\rm{s}}/R_{cf}$,  with $d_{\rm{s}}$ the shock 'stand-off' distance and $R_{\rm{cf}}$ the radius of curvature of the cold front, depends on ${\cal M}$ as a function of $({\cal M}^2-1)^{-1}$ (see Dasadia et al. Fig. 9, \citealt{Schreier1982,Vikhlinin2001,Sarazin2001}). For a shock with ${\cal M}_E$~=~1.5~$\pm$~0.1, the shock offset ratio is $d_{\rm{s}}/R_{\rm{cf}}$~$\sim$~0.8~$\pm$~0.1, $d_{\rm{s}}$~$\sim$~120~$\pm$~15 kpc ($R_{\rm{cf}}$~$\sim$~150 kpc). This separation for a rigid sphere is smaller than the distance of $\sim$~300 kpc between the cold front  and shock front present in the East. This behaviour is the same for most of the observed clusters  (\citealt{Dasadia2016b}). As explained by \cite{Sarazin2016}, after the first core passage the shock accelerates towards the periphery of the clusters, while the cool gas is decelerated by gravity and ram pressure. For this reason, the shock is expected to move away from the cold front and the separation between both features migh be larger than the rigid sphere model predicts.

\section{Summary}

We present a spectral analysis of A3376 in four directions (West, East, North and South) using \textit{Suzaku} deep observations and supported by \textit{XMM-Newton} and \textit{Chandra} observations. A3376 is a  merging galaxy cluster with two shock fronts and one cold front confirmed. One shock is coincident with the radio relic in the West and the other might be associated to the 'notch' of the eastern relic. Shocks are moving with a Mach number ${\cal M}$ $\sim$ 2--3. The cold front is located at approximately 150 kpc ($r\sim~3$\arcmin\ from the X-ray emission peak at the center and delimits a cool gas cloud moving with $v$~$\sim$~1300 km s$^{-1}$.

We determine the ICM structure up to 0.9$r_{200}$ in the W, 0.6$r_{200}$ in the E. We observe a temperature enhancement followed by a drop at $\sim$0.7$r_{200}$ for the W and  $\sim$0.5$r_{200}$ for the E. We confirm temperature and surface brightness discontinuities in both directions. In the West it coincides with the radio relic position and in the East, it is located at 450 kpc ($r\sim~8$\arcmin). The temperature structure in the South follows the simulated profile of \cite{Burns2010} for relaxed clusters.

We estimate the Mach number based on the temperature jump, being ${\cal M}_{\rm{W}}$ = 2.8~$\pm~0.4$ and  ${\cal M}_{\rm{E}}$ = 1.5~$\pm~0.1$, for the western and eastern region, respectively. We derive the shock speed velocity as $v_{\rm{shock,W}}$~$\sim$~1630~$\pm~220$ km s$^{-1}$ and  $v_{\rm{shock,E}}$~$\sim$~1450~$\pm~150$ km s$^{-1}$.   

Assuming that the shock fronts are moving with constant velocity, the time since core passage is $\sim$0.6 Gyr, which is in good agreement with the N-body hydrodynamical simulations of \cite{Machado2013}, and with earlier multiwavelength analyses of  \cite{George2015} and \cite{Monteiro-Oliveira2017}. Combining the eastern shock velocity and the peculiar velocity of the eastern brightest galaxy let us infer that the merger axis is close to the plane of the sky.

 We compare the Mach numbers inferred from the temperature discontinuities with those  derived from radio observations assuming the DSA mechanism. Not all the clusters present a consistent correlation between X-ray and radio Mach numbers. For A3376 the Mach number for the western relic estimated by \cite{Kale2012} is consistent with our result.


\begin{acknowledgements}
The authors thank the anonymous referee for constructive comments and suggestions. The authors thank Dr. R. Kale for providing the VLA radio data. I.U. thanks H. Mart{\'{\i}}nez-Rodr{\'{\i}}guez for his support in the initial stage of this work.  H.A. acknowledges the support of NWO via Veni grant. T.O. and Y.I. acknowledge support by JSPS KAKENHI Grant Numbers 26220703 and 15H03642. SRON is supported
financially by NWO, the Netherlands Organization for Scientific
Research. This work is based on observations obtained with XMM-Newton, an ESA science mission with instruments and contributions directly funded by ESA member states and the USA (NASA). 
\end{acknowledgements}


\begin{thebibliography}{99}
 
\bibitem[{Akamatsu {et~al.}(2013)Akamatsu, Inoue, Sato, Matsusita, Ishisaki, \&
  Sarazin}]{Akamatsu2013b}
Akamatsu, H., Inoue, S., Sato, T., {et~al.} 2013, PASJ, 65, 89

\bibitem[{Akamatsu \& Kawahara(2013)}]{Akamatsu2013a}
Akamatsu, H. \& Kawahara, H. 2013, PASJ, 65, 16

\bibitem[{Akamatsu {et~al.}(2017)Akamatsu, Mizuno, Ota, Zhang, van Weeren,
  Kawahara, Fukazawa, Kaastra, Kawaharada, Nakazawa, Ohashi, R{\"{o}}ttgering,
  Takizawa, Vink, \& Zandanel}]{Akamatsu2017}
Akamatsu, H., Mizuno, M., Ota, N., {et~al.} 2017, A{\&}A, 600, A100

\bibitem[{Akamatsu {et~al.}(2012)Akamatsu, Takizawa, Nakazawa, Fukazawa,
  Ishisaki, \& Ohashi}]{Akamatsu2012b}
Akamatsu, H., Takizawa, M., Nakazawa, K., {et~al.} 2012, PASJ, 64

\bibitem[{Akamatsu {et~al.}(2015)Akamatsu, van Weeren, Ogrean, Kawahara, Stroe,
  Sobral, Hoeft, R{\"{o}}ttgering, Br{\"{u}}ggen, \& Kaastra}]{Akamatsu2015}
Akamatsu, H., van Weeren, R.~J., Ogrean, G.~A., {et~al.} 2015, A{\&}A, 582, A20

\bibitem[{Bagchi {et~al.}(2006)Bagchi, Durret, {Lima Neto}, \&
  Paul}]{Bagchi2006}
Bagchi, J., Durret, F., {Lima Neto}, G.~B., \& Paul, S. 2006, Science, 314, 791

\bibitem[{Bell(1987)}]{Bell1987}
Bell, A.~R. 1987, MNRAS, 225, 615

\bibitem[{Blandford \& Eichler(1987)}]{Blandford1987}
Blandford, R. \& Eichler, D. 1987, Phys. Rep., 154, 1

\bibitem[{{Blanton} {et~al.}(2003){Blanton}, {Sarazin}, \&
  {McNamara}}]{Blanton2003}
{Blanton}, E.~L., {Sarazin}, C.~L., \& {McNamara}, B.~R. 2003, \apj, 585, 227

\bibitem[{Bonafede {et~al.}(2014)Bonafede, Intena, Br{\"{u}}ggen, Girardi,
  Nonino, Kantharia, van Weeren, \& R{\"{o}}ttgering}]{Bonafede2014}
Bonafede, A., Intena, H.~T., Br{\"{u}}ggen, M., {et~al.} 2014, ApJ, 785, 1

\bibitem[{Botteon {et~al.}(2016a)Botteon, Gastaldello, Brunetti, \&
  Dallacasa}]{Botteon2016a}
Botteon, A., Gastaldello, F., Brunetti, G., \& Dallacasa, D. 2016a, MNRAS, 460,
  84

\bibitem[{Botteon {et~al.}(2016b)Botteon, Gastaldello, Brunetti, \&
  Kale}]{Botteon2016b}
Botteon, A., Gastaldello, F., Brunetti, G., \& Kale, R. 2016b, MNRAS, 463, 1534

\bibitem[{Bourdin {et~al.}(2013)Bourdin, Mazzotta, Markevitch, Giacintucci, \&
  Brunetti}]{Bourdin2013}
Bourdin, H., Mazzotta, P., Markevitch, M., Giacintucci, S., \& Brunetti, G.
  2013, ApJ, 764, 82

\bibitem[{Br{\"{u}}ggen {et~al.}(2012)Br{\"{u}}ggen, Bykov, Ryu, \&
  R{\"{o}}ttgering}]{Bruggen2012}
Br{\"{u}}ggen, M., Bykov, A., Ryu, D., \& R{\"{o}}ttgering, H. 2012, Space Sci.
  Rev., 166, 187

\bibitem[{Brunetti \& Jones(2014)}]{Brunetti2014}
Brunetti, G. \& Jones, T.~W. 2014, IJMPD, 23, 1430007

\bibitem[{Burns {et~al.}(2010)Burns, Skillman, \& O'Shea}]{Burns2010}
Burns, J.~O., Skillman, S.~W., \& O'Shea, B.~W. 2010, ApJ, 721, 1105

\bibitem[{Cappelluti {et~al.}(2017)Cappelluti, Li, Ricarte, Agarwal, Allevato,
  Ananna, Ajello, Civano, Comastri, Elvis, Finoguenov, Gilli, Hasinger,
  Marchesi, Natarajan, Pacucci, Treister, \& Urry}]{Cappelluti2017}
Cappelluti, N., Li, Y., Ricarte, A., {et~al.} 2017, ApJ, 837, 19

\bibitem[{Cash(1979)}]{Cash1979}
Cash, W. 1979, ApJ, 228, 939

\bibitem[{Dasadia {et~al.}(2016b)Dasadia, Sun, Morandi, Sarazin, Clarke,
  Nulsen, Massaro, Roediger, Harris, \& Forman}]{Dasadia2016b}
Dasadia, S., Sun, M., Morandi, A., {et~al.} 2016b, MNRAS, 458, 681

\bibitem[{Dasadia {et~al.}(2016a)Dasadia, Sun, Sarazin, Morandi, Markevitch,
  Wik, Feretti, Giovannini, Govoni, \& Vacca}]{Dasadia2016a}
Dasadia, S., Sun, M., Sarazin, C., {et~al.} 2016a, ApJ, 820, L20

\bibitem[{{De Grandi} \& Molendi(2002)}]{DeGrandi2002}
{De Grandi}, S. \& Molendi, S. 2002, ApJ, 567, 163

\bibitem[{Drury(1983)}]{Drury1983}
Drury, L.~O. 1983, Reports on Progress in Physics, 46, 973

\bibitem[{Durret {et~al.}(2013)Durret, Perrot, {Lima Neto}, Adami, Bertin, \&
  Bagchi}]{Durret2013}
Durret, F., Perrot, C., {Lima Neto}, G.~B., {et~al.} 2013, A{\&}A, 560, A78

\bibitem[{{Eckert} {et~al.}(2016a){Eckert}, {Ettori}, {Coupon}, {Gastaldello},
  {Pierre}, {Melin}, {Le Brun}, {McCarthy}, {Adami}, {Chiappetti}, {Faccioli},
  {Giles}, {Lavoie}, {Lef{\`e}vre}, {Lieu}, {Mantz}, {Maughan}, {McGee},
  {Pacaud}, {Paltani}, {Sadibekova}, {Smith}, \& {Ziparo}}]{Eckert2016a}
{Eckert}, D., {Ettori}, S., {Coupon}, J., {et~al.} 2016a, \aap, 592, A12

\bibitem[{Eckert {et~al.}(2016b)Eckert, Jauzac, Vazza, Owers, Kneib, Tchernin,
  Intema, \& Knowles}]{Eckert2016b}
Eckert, D., Jauzac, M., Vazza, F., {et~al.} 2016b, MNRAS, 461, 1302

\bibitem[{Eckert {et~al.}(2011)Eckert, Molendi, \& Paltani}]{Eckert2011}
Eckert, D., Molendi, S., \& Paltani, S. 2011, A{\&}A, 526, A79

\bibitem[{Feretti {et~al.}(2012)Feretti, Giovannini, Govoni, \&
  Murgia}]{Feretti2012}
Feretti, L., Giovannini, G., Govoni, F., \& Murgia, M. 2012, A{\&}A Rev., 20,
  54

\bibitem[{Ferrari {et~al.}(2008)Ferrari, Govoni, Schindler, Bykov, \&
  Rephaeli}]{Ferrari2008}
Ferrari, C., Govoni, F., Schindler, S., Bykov, A.~M., \& Rephaeli, Y. 2008,
  Space Sc. Rev., 134, 93

\bibitem[{Finoguenov {et~al.}(2010)Finoguenov, Sarazin, Nakazawa, Wik, \&
  Clarke}]{Finoguenov2010}
Finoguenov, A., Sarazin, C.~L., Nakazawa, K., Wik, D.~R., \& Clarke, T.~E.
  2010, ApJ, 715, 1143

\bibitem[{{Fujita} {et~al.}(2016){Fujita}, {Akamatsu}, \&
  {Kimura}}]{Fujita2016}
{Fujita}, Y., {Akamatsu}, H., \& {Kimura}, S.~S. 2016, \pasj, 68, 34

\bibitem[{Fujita {et~al.}(2015)Fujita, Takizawa, Yamazaki, Akamatsu, \&
  Ohno}]{Fujita2015}
Fujita, Y., Takizawa, M., Yamazaki, R., Akamatsu, H., \& Ohno, H. 2015, ApJ,
  815, 116

\bibitem[{Fujita {et~al.}(2008)Fujita, Tawa, Hayashida, Takizawa, Matsumoto,
  Okabe, \& Reiprich}]{Fujita2008}
Fujita, Y., Tawa, N., Hayashida, K., {et~al.} 2008, PASJ, 60, 343

\bibitem[{George {et~al.}(2015)George, Dwarakanath, Johnston-Hollitt,
  Hurley-Walker, Hindson, Kapiska, Tingay, Bell, Callingham, For, Hancock,
  Lenc, Mckinley, Morgan, Offringa, Procopio, Staveley-Smith, Wayth, Wu, Zheng,
  Bernardi, Kaplan, Kasper, Kratzenberg, Lonsdale, Lynch, Mcwhirter, Mitchell,
  Morales, Morgan, Oberoi, Ord, Prabu, Rogers, Roshi, Shankar, Srivani,
  Subrahmanyan, Waterson, Webster, Whitney, Williams, \& Williams}]{George2015}
George, L.~T., Dwarakanath, K.~S., Johnston-Hollitt, M., {et~al.} 2015, MNRAS,
  451, 4207

\bibitem[{Giacintucci {et~al.}(2008)Giacintucci, Venturi, Macario, Dallacasa,
  Brunetti, Markevitch, Cassano, Bardelli, \& Athreya}]{Giacintucci2008}
Giacintucci, S., Venturi, T., Macario, G., {et~al.} 2008, A{\&}A, 486, 347

\bibitem[{{Guo} {et~al.}(2014{\natexlab{a}}){Guo}, {Sironi}, \&
  {Narayan}}]{Guo2014a}
{Guo}, X., {Sironi}, L., \& {Narayan}, R. 2014{\natexlab{a}}, \apj, 794, 153

\bibitem[{{Guo} {et~al.}(2014{\natexlab{b}}){Guo}, {Sironi}, \&
  {Narayan}}]{Guo2014b}
{Guo}, X., {Sironi}, L., \& {Narayan}, R. 2014{\natexlab{b}}, \apj, 797, 47

\bibitem[{{Guo} {et~al.}(2017){Guo}, {Sironi}, \& {Narayan}}]{Guo2017}
{Guo}, X., {Sironi}, L., \& {Narayan}, R. 2017, \apj, 851, 134

\bibitem[{Henry {et~al.}(2009)Henry, Evrard, Hoekstra, Babul, \&
  Mahdavi}]{Henry2009}
Henry, J.~P., Evrard, A.~E., Hoekstra, H., Babul, A., \& Mahdavi, A. 2009, ApJ,
  691, 1307

\bibitem[{Hindson {et~al.}(2014)Hindson, Johnston-Hollitt, Hurley-Walker,
  Buckley, Morgan, Carretti, Dwarakanath, Bell, Bernardi, Bhat, Bowman, Briggs,
  Cappallo, Corey, Deshpande, Emrich, Ewall-Wice, Feng, Gaensler, Goeke,
  Greenhill, Hazelton, Jacobs, Kaplan, Kasper, Kratzenberg, Kudryavtseva, Lenc,
  Lonsdale, Lynch, McWhirter, McKinley, Mitchell, Morales, Morgan, Oberoi, Ord,
  Pindor, Prabu, Procopio, Offringa, Riding, Rogers, Roshi, {Udaya Shankar},
  Srivani, Subrahmanyan, Tingay, Waterson, Wayth, Webster, Whitney, Williams,
  \& Williams}]{Hindson2014}
Hindson, L., Johnston-Hollitt, M., Hurley-Walker, N., {et~al.} 2014, MNRAS,
  445, 330

\bibitem[{Hoang {et~al.}(2017)Hoang, Shimwell, Stroe, Akamatsu, Brunetti,
  Donnert, Intema, Mulcahy, R{\"{o}}ttgering, van Weeren, Bonafede,
  Br{\"{u}}ggen, Cassano, Chy{\.{z}}y, En{\ss}lin, Ferrari, de~Gasperin, Gu,
  Hoeft, Miley, Orr{\'{u}}, Pizzo, \& White}]{Hoang2017}
Hoang, D.~N., Shimwell, T.~W., Stroe, A., {et~al.} 2017, MNRAS, 471, 1107

\bibitem[{Hoshino {et~al.}(2010)Hoshino, Henry, Sato, Akamatsu, Yokota, Sasaki,
  Ishisaki, Ohashi, Bautz, Fukazawa, Kawano, Furuzawa, Hayashida, Tawa, Hughes,
  Kokubun, \& Tamura}]{Hoshino2010}
Hoshino, A., Henry, J.~P., Sato, K., {et~al.} 2010, PASJ, 62, 371

\bibitem[{Ichinohe {et~al.}(2017)Ichinohe, Simionescu, Werner, \&
  Takahashi}]{Ichinohe2017}
Ichinohe, Y., Simionescu, A., Werner, N., \& Takahashi, T. 2017, MNRAS, 467,
  3662

\bibitem[{Ishisaki {et~al.}(2007)Ishisaki, Maeda, Fujimoto, Ozaki, Ebisawa,
  Takahashi, Ueda, Ogasaka, Ptak, Mukai, Hamaguchi, Hirayama, Otani, Ubo, \&
  Hibata}]{Ishisaki2007}
Ishisaki, Y., Maeda, Y., Fujimoto, R., {et~al.} 2007, PASJ, 59, 113

\bibitem[{Itahana {et~al.}(2015)Itahana, Takizawa, Akamatsu, Ohashi, Ishisaki,
  Kawahara, \& {Van Weeren}}]{Itahana2015}
Itahana, M., Takizawa, M., Akamatsu, H., {et~al.} 2015, PASJ, 67, 1

\bibitem[{Kaastra(2017)}]{Kaastra2017}
Kaastra, J.~S. 2017, A{\&}A, 605, A51

\bibitem[{Kaastra \& Bleeker(2016)}]{Kaastra2016}
Kaastra, J.~S. \& Bleeker, J. A.~M. 2016, A{\&}A, 587, A151

\bibitem[{{Kaastra} {et~al.}(1996){Kaastra}, {Mewe}, \&
  {Nieuwenhuijzen}}]{Kaastra1996}
{Kaastra}, J.~S., {Mewe}, R., \& {Nieuwenhuijzen}, H. 1996, in UV and X-ray
  Spectroscopy of Astrophysical and Laboratory Plasmas, ed. K.~{Yamashita} \&
  T.~{Watanabe}, 411--414

\bibitem[{Kale {et~al.}(2012)Kale, Dwarakanath, Bagchi, \& Paul}]{Kale2012}
Kale, R., Dwarakanath, K.~S., Bagchi, J., \& Paul, S. 2012, MNRAS, 426, 1204

\bibitem[{Kang(2017)}]{Kang2017}
Kang, H. 2017, JKAS, 50, 93

\bibitem[{Kang \& Ryu(2015)}]{Kang2015}
Kang, H. \& Ryu, D. 2015, ApJ, 809, 186

\bibitem[{Kang \& Ryu(2016)}]{kang2016}
Kang, H. \& Ryu, D. 2016, ApJ, 823, 13

\bibitem[{Kang {et~al.}(2012)Kang, Ryu, \& Jones}]{Kang2012}
Kang, H., Ryu, D., \& Jones, T.~W. 2012, ApJ, 756, 97

\bibitem[{Kawano {et~al.}(2009)Kawano, Fukazawa, Nishino, Nakazawa, Kitaguchi,
  Makishima, Takahashi, Kokubun, Ota, Ohashi, Isobe, Henry, \&
  Hornschemeier}]{Kawano2008}
Kawano, N., Fukazawa, Y., Nishino, S., {et~al.} 2009, PASJ, 61, 377

\bibitem[{Koyama {et~al.}(2007)Koyama, Tsunemi, Dotani, Bautz, Hayashida,
  Tsuru, Matsumoto, Ogawara, Ricker, Doty, Kissel, Foster, Nakajima, Yamaguchi,
  Mori, Sakano, Hamaguchi, Nishiuchi, Miyata, Torii, Namiki, Katsuda, Matsuura,
  Miyauchi, Anabuki, Tawa, Ozaki, Murakami, Maeda, Ichikawa, Prigozhin,
  Boughan, LaMarr, Miller, Burke, Gregory, Pillsbury, Bamba, Hiraga, Senda,
  Katayama, Kitamoto, Tsujimoto, Kohmura, Tsuboi, \& Awaki}]{Koyama2007}
Koyama, K., Tsunemi, H., Dotani, T., {et~al.} 2007, PASJ, 59, S23

\bibitem[{Kushino {et~al.}(2002)Kushino, Ishisaki, Morita, Yamasaki, Ishida,
  Ohashi, \& Ueda}]{Kushino2002}
Kushino, A., Ishisaki, Y., Morita, U., {et~al.} 2002, PASJ, 54, 327

\bibitem[{Landau \& Lifshitz(1959)}]{Landau1959}
Landau, L.~D. \& Lifshitz, E.~M. 1959, Fluid mechanics, Course of theoretical
  physics, Oxford: Pergamon Press

\bibitem[{Lindner {et~al.}(2014)Lindner, Baker, Hughes, Battaglia, Gupta,
  Knowles, Marriage, Menanteau, Moodley, Reese, \& Srianand}]{Lindner2014}
Lindner, R.~R., Baker, A.~J., Hughes, J.~P., {et~al.} 2014, ApJ, 786, 49

\bibitem[{Lodders {et~al.}(2009)Lodders, Palme, \& Gail}]{Lodders2009}
Lodders, K., Palme, H., \& Gail, H.-P. 2009, Landolt B{\"{o}}mstein, New
  Series, Astronomy and Astrophysics, Springer Verlag,, VI/4B, 560

\bibitem[{Macario {et~al.}(2011)Macario, Markevitch, Giacintucci, Brunetti,
  Venturi, \& Murray}]{Macario2011}
Macario, G., Markevitch, M., Giacintucci, S., {et~al.} 2011, ApJ, 782, 82

\bibitem[{Machado \& {Lima Neto}(2013)}]{Machado2013}
Machado, R. E.~G. \& {Lima Neto}, G.~B. 2013, MNRAS, 430, 3249

\bibitem[{Markevitch {et~al.}(2002)Markevitch, Gonzalez, David, Vikhlinin,
  Murray, Forman, Jones, \& Tucker}]{Markevitch2002}
Markevitch, M., Gonzalez, A.~H., David, L., {et~al.} 2002, ApJ, 567, L27

\bibitem[{Markevitch {et~al.}(2005)Markevitch, Govoni, Brunetti, \&
  Jerius}]{Markevitch2005}
Markevitch, M., Govoni, F., Brunetti, G., \& Jerius, D. 2005, ApJ, 627, 733

\bibitem[{Markevitch \& Vikhlinin(2007)}]{Markevitch2007}
Markevitch, M. \& Vikhlinin, A. 2007, Phys. Rep., 443, 1

\bibitem[{Mazzota {et~al.}(2011)Mazzota, Bourdin, Giacintucci, Markevitch, \&
  Venturi}]{Mazzota2011}
Mazzota, P., Bourdin, H., Giacintucci, S., Markevitch, M., \& Venturi, T. 2011,
  Mem. Soc. Astron. Italiana, 82, 495

\bibitem[{Miniati(2002)}]{Miniati2002}
Miniati, F. 2002, MNRAS, 337, 199

\bibitem[{Mitsuda {et~al.}(2007)Mitsuda, Bautz, Inoue, Kelley, Koyama, Kunieda,
  \& Makishima}]{Mitsuda2007}
Mitsuda, K., Bautz, M., Inoue, H., {et~al.} 2007, PASJ, 59, 1

\bibitem[{Monteiro-Oliveira {et~al.}(2017)Monteiro-Oliveira, Neto, Cypriano,
  Machado, Capelato, Lagan{\'{a}}, Durret, \& Bagchi}]{Monteiro-Oliveira2017}
Monteiro-Oliveira, R., Neto, G. B.~L., Cypriano, E.~S., {et~al.} 2017, MNRAS,
  468, 4566

\bibitem[{Ogrean \& Br{\"{u}}ggen(2013)}]{Ogrean2013}
Ogrean, G.~A. \& Br{\"{u}}ggen, M. 2013, MNRAS, 433, 1701

\bibitem[{Ogrean {et~al.}(2014)Ogrean, Br{\"{u}}ggen, van Weeren, Burgmeier, \&
  Simionescu}]{Ogrean2014}
Ogrean, G.~A., Br{\"{u}}ggen, M., van Weeren, R.~J., Burgmeier, A., \&
  Simionescu, A. 2014, MNRAS, 443, 2463

\bibitem[{Owers {et~al.}(2014)Owers, Nulsen, Couch, Ma, David, Forman, Hopkins,
  Jones, \& van Weeren}]{Owers2014}
Owers, M.~S., Nulsen, P. E.~J., Couch, W.~J., {et~al.} 2014, ApJ, 780, 163

\bibitem[{Pacholczyk(1973)}]{Pacholczyk1970}
Pacholczyk, A.~G. 1973, Radio astrophysics. Nonthermal processes in galactic
  and extragalactic sources, Moskva: Mir

\bibitem[{Pinzke {et~al.}(2013)Pinzke, Oh, \& Pfrommer}]{Pinzke2013}
Pinzke, A., Oh, S.~P., \& Pfrommer, C. 2013, MNRAS, 435, 1061

\bibitem[{Pizzo \& de~Bruyn(2009)}]{Pizzo2009}
Pizzo, R.~F. \& de~Bruyn, a.~G. 2009, A{\&}A, 507, 639

\bibitem[{Reiprich {et~al.}(2013)Reiprich, Basu, Ettori, Israel, Lovisari,
  Molendi, Pointecouteau, \& Roncarelli}]{Reiprich2013}
Reiprich, T.~H., Basu, K., Ettori, S., {et~al.} 2013, Space Sci. Rev., 177, 195

\bibitem[{Reiprich {et~al.}(2009)Reiprich, Hudson, Zhang, Sato, Ishisaki,
  Hoshino, Ohashi, Ota, \& Fujita}]{Reiprich2009}
Reiprich, T.~H., Hudson, D.~S., Zhang, Y.~Y., {et~al.} 2009, A{\&}A, 501, 899

\bibitem[{Russell {et~al.}(2014)Russell, Fabian, McNamara, Edge, Sanders,
  Nulsen, Baum, Donahue, \& O'Dea}]{Russell2014}
Russell, H.~R., Fabian, A.~C., McNamara, B.~R., {et~al.} 2014, MNRAS, 444, 629

\bibitem[{Russell {et~al.}(2012)Russell, Mcnamara, Sanders, Fabian, Nulsen,
  Canning, Baum, Donahue, Edge, King, \& O'Dea}]{Russell2012}
Russell, H.~R., Mcnamara, B.~R., Sanders, J.~S., {et~al.} 2012, MNRAS, 423, 236

\bibitem[{Sarazin(2001)}]{Sarazin2001}
Sarazin, C.~L. 2001, Merging Processes in Clusters of Galaxies, Astrophysics
  and Space Science Library, Vol. 272. Kluwer Academic Publishers, Dordrecht, 1

\bibitem[{Sarazin {et~al.}(2016)Sarazin, Finoguenov, Wik, \&
  Clarke}]{Sarazin2016}
Sarazin, C.~L., Finoguenov, A., Wik, D.~R., \& Clarke, T.~E. 2016, ArXiv
  e-prints [arXiv:1606.07433]

\bibitem[{Schreier(1982)}]{Schreier1982}
Schreier, S. 1982, in Compressible Flow.Wiley, New York

\bibitem[{Serlemitsos {et~al.}(2007)Serlemitsos, Soong, Chan, Okajima, Lehan,
  Maeda, Itoh, Mori, Iizuka, Itoh, Inoue, Okada, Yokoyama, Itoh, Ebara,
  Nakamura, Suzuki, Ishida, Hayakawa, Inoue, Okuma, Kubota, Suzuki, Osawa,
  Yamashita, Kunieda, Tawara, Ogasaka, Furuzawa, Tamura, Shibata, Haba, Naitou,
  \& Misaki}]{Serlemitsos2007}
Serlemitsos, P.~J., Soong, Y., Chan, K.~W., {et~al.} 2007, PASJ, 59, 9

\bibitem[{Shimwell {et~al.}(2016)Shimwell, Luckin, Br, Brunetti, Intema, Owers,
  Stroe, Weeren, Williams, Cassano, Gasperin, Heald, Hoang, Hardcastle,
  Sridhar, Sabater, Best, Bonafede, Chy, Ferrari, Haverkorn, Hoeft, Horellou,
  \& Mckean}]{Shimwell2016}
Shimwell, T.~W., Luckin, J., Br, M., {et~al.} 2016, MNRAS, 459, 277

\bibitem[{{Shimwell} {et~al.}(2015){Shimwell}, {Markevitch}, {Brown},
  {Feretti}, {Gaensler}, {Johnston-Hollitt}, {Lage}, \&
  {Srinivasan}}]{Shimwell2015}
{Shimwell}, T.~W., {Markevitch}, M., {Brown}, S., {et~al.} 2015, \mnras, 449,
  1486

\bibitem[{Skillman {et~al.}(2013)Skillman, Xu, Hallman, O'Shea, Burns, Li,
  Collins, \& Norman}]{Skillman2013}
Skillman, S.~W., Xu, H., Hallman, E.~J., {et~al.} 2013, ApJ, 765

\bibitem[{Smith {et~al.}(2004)Smith, Hudson, Nelan, Moore, Quinney, Wegner,
  Lucey, Davies, Malecki, Schade, \& Suntzeff}]{Smith2004}
Smith, R.~J., Hudson, M.~J., Nelan, J.~E., {et~al.} 2004, ApJ, 128, 1558

\bibitem[{{Storm} {et~al.}(2017){Storm}, {Vink}, {Zandanel}, \&
  {Akamatsu}}]{Storm2017}
{Storm}, E., {Vink}, J., {Zandanel}, F., \& {Akamatsu}, H. 2017, ArXiv e-prints [arXiv:1712.04539]

\bibitem[{Stroe {et~al.}(2014)Stroe, Harwood, Hardcastle, \&
  R{\"{o}}ttgering}]{Stroe2014}
Stroe, A., Harwood, J.~J., Hardcastle, M.~J., \& R{\"{o}}ttgering, H.~J. 2014,
  MNRAS, 445, 1213

\bibitem[{Struble \& Rood(1999)}]{Struble1999}
Struble, M.~F. \& Rood, H.~J. 1999, ApJS, 125, 35

\bibitem[{Tawa {et~al.}(2008)Tawa, Hayashida, Nagai, Nakamoto, Tsunemi,
  Yamaguchi, Ishisaki, Miller, Mizuno, Dotani, Ozaki, \& Katayama}]{Tawa2008}
Tawa, N., Hayashida, K., Nagai, M., {et~al.} 2008, PASJ, 60, 11

\bibitem[{Thierbach {et~al.}(2003)Thierbach, Klein, \&
  Wielebinski}]{Thierbach2003}
Thierbach, M., Klein, U., \& Wielebinski, R. 2003, A{\&}A, 397, 53

\bibitem[{Trasatti {et~al.}(2015)Trasatti, Akamatsu, Lovisari, Klein, Bonafede,
  Bruggen, Dallacasa, \& Clarke}]{Trasatti2015}
Trasatti, M., Akamatsu, H., Lovisari, L., {et~al.} 2015, A{\&}A, 575, A45

\bibitem[{Urban {et~al.}(2017)Urban, Werner, Allen, Simionescu, \&
  Mantz}]{Urban2017}
Urban, O., Werner, N., Allen, S.~W., Simionescu, A., \& Mantz, A. 2017, MNRAS,
  470, 4583

\bibitem[{van Weeren {et~al.}(2017)van Weeren, Andrade-Santos, Dawson,
  Golovich, Lal, Kang, Ryu, Br{\`{i}}ggen, Ogrean, Forman, Jones, Placco,
  Santucci, Wittman, Jee, Kraft, Sobral, Stroe, \& Fogarty}]{VanWeeren2017}
van Weeren, R.~J., Andrade-Santos, F., Dawson, W.~A., {et~al.} 2017, Nature
  Astronomy, 1, 5

\bibitem[{van Weeren {et~al.}(2016)van Weeren, Brunetti, Br{\"{u}}ggen,
  Andrade-Santos, Ogrean, Williams, R{\"{o}}ttgering, Dawson, Forman,
  de~Gasperin, Hardcastle, Jones, Miley, Rafferty, Rudnick, Sabater, Sarazin,
  Shimwell, Bonafede, Best, B{\^\i}rzan, Cassano, Chy{\.{z}}y, Croston,
  Dijkema, En{\ss}lin, Ferrari, Heald, Hoeft, Horellou, Jarvis, Kraft, Mevius,
  Intema, Murray, Orr{\'{u}}, Pizzo, Sridhar, Simionescu, Stroe, van~der Tol,
  \& White}]{VanWeeren2016}
van Weeren, R.~J., Brunetti, G., Br{\"{u}}ggen, M., {et~al.} 2016, ApJ, 818,
  204

\bibitem[{van Weeren {et~al.}(2012)van Weeren, R{\"{o}}ttgering, Rafferty,
  Pizzo, Bonafede, Br{\"{u}}ggen, Brunetti, Ferrari, Orr{\`{u}}, Heald, McKean,
  Tasse, de~Gasperin, B{\^{i}}rzan, van Zwieten, van~der Tol, Shulevski,
  Jackson, Offringa, Conway, Intema, Clarke, van Bemmel, Miley, White, Hoeft,
  Cassano, Macario, Morganti, Wise, Horellou, Valentijn, Wucknitz, Kuijken,
  En{\ss}lin, Anderson, Asgekar, Avruch, Beck, Bell, Bell, Bentum, Bernardi,
  Best, Boonstra, Brentjens, van~de Brink, Broderick, Brouw, Butcher, van
  Cappellen, Ciardi, Eisl{\"{o}}ffel, Falcke, Fender, Garrett, Gerbers, Gunst,
  van Haarlem, Hamaker, Hassall, Hessels, Koopmans, Kuper, van Leeuwen, Maat,
  Millenaar, Munk, Nijboer, Noordam, Pandey, Pandey-Pommier, Polatidis, Reich,
  Scaife, Schoenmakers, Sluman, Stappers, Steinmetz, Swinbank, Tagger, Tang,
  Vermeulen, de~Vos, \& van Haarlem}]{VanWeeren2012}
van Weeren, R.~J., R{\"{o}}ttgering, H. J.~A., Rafferty, D.~A., {et~al.} 2012,
  A{\&}A, 543, A43

\bibitem[{Vazza \& Br{\"{u}}ggen(2014)}]{Vazza2014}
Vazza, F. \& Br{\"{u}}ggen, M. 2014, MNRAS, 437, 2291

\bibitem[{Vikhlinin {et~al.}(2001)Vikhlinin, Markevitch, \&
  Murray}]{Vikhlinin2001}
Vikhlinin, A., Markevitch, M., \& Murray, S.~S. 2001, ApJ, 551, 160

\bibitem[{{Vink} \& {Yamazaki}(2014)}]{Vink2014}
{Vink}, J. \& {Yamazaki}, R. 2014, \apj, 780, 125

\bibitem[{Willingale {et~al.}(2013)Willingale, Starling, Beardmore, Tanvir, \&
  O'Brien}]{Willingale2013}
Willingale, R., Starling, R. L.~C., Beardmore, A.~P., Tanvir, N.~R., \&
  O'Brien, P.~T. 2013, MNRAS, 431, 394

\end{thebibliography}


\end{document}